\documentclass[preprint,prb,showpacs,superscriptaddress]{revtex4-1}
\usepackage{amsmath}
\usepackage{graphicx}
\usepackage{amssymb}
\usepackage{times}
\usepackage{color}
\usepackage{multirow}
\definecolor{lr}{rgb}{1.0,0.3,0.3}
\definecolor{dg}{rgb}{0.0,0.5,0.0}

\usepackage{comment}

\graphicspath{./}
 


\begin{document}

\title{Longitudinal spin relaxation model applied to point defect qubit systems}

\author{Viktor Iv\'ady}
\email{ivady.viktor@wigner.hu}
\affiliation{Wigner Research Centre for Physics
  PO Box 49, H-1525, Budapest, Hungary}
\affiliation{Department of Physics, Chemistry and Biology, Link\"oping
  University, SE-581 83 Link\"oping, Sweden}

\date{\today}

\begin{abstract}
Controllable, partially isolated few level systems in semiconductors have recently gained multidisciplinary attention due to their widespread nanoscale sensing and quantum technology applications.  Quantitative simulation of the dynamics and related applications of such systems is a challenging theoretical task that requires faithful description not only the few level systems but also their local environments. Here, we develop a method that can describe relevant relaxation processes induced by a dilute bath of nuclear and electron spins. The method utilizes an extended Lindblad equation in the framework of cluster approximation of a central spin system. We demonstrate that the proposed method can accurately describe T$_1$ time of an exemplary solid-state point defect qubit system, in particular  NV center in diamond, at various magnetic fields and strain. 
\end{abstract}
\maketitle


\section{Introduction}

Controllable solid-state spin systems have attracted considerable scientific and technological interest over the last decades. Point defect-based applications are among the most recent use of solid-state spins that allow full control over a set of electron and nuclear spins. The NV center, substitution nitrogen-carbon vacancy complex point defect in negative charge state in diamond\citep{duPreez:1965,Wrachtrup:JPCM2006,Maze2011,DOHERTY20131,GaliReview2019} is a magneto-optically active electron spin system that can be isolated to a large degree from the environmental disturbances. The NV center's triplet electron spin can be initialized by pumping through optically excited triplet and meta stable singlet states.\citep{Maze2011} The very same process gives rise to spin dependent optical decay that allows high fidelity read-out even at single NV center level.\citep{Gruber:Science1997,Jelezko:PRL200492,Siyushev728}  In association with nuclear spins, NV center can implement few qubit nodes to realize high fidelity gates.\citep{Childress:Science2006} Coherence time may exceed a millisecond\citep{Balasubramanian:NatMat2009} and the qubit nodes can operate even above 600~K.\citep{ToyliPhysRevX2012} These attributes made NV center interesting for a broad range of quantum technology applications, especially in the field of quantum sensing\citep{Maze:Nature2008,Dolde2011,Kucsko2013,Teissier2014} and quantum information processing\citep{Wrachtrup:JPCM2006,Weber10,Awschalom2013}. Besides NV center, there have been several akin point defect qubit systems demonstrated in various wide band gap semiconductors.\citep{Koehl11,Widmann2015,Rose2018}

Environmental spins, such as point defect and nuclear spins, play a crucial role in spin relaxation and decoherence processes that are often the major limiting factors in quantum technology applications. Due to the complexity of some environmental spins' inner energy level structure, decay processes often depend on external control parameters, such as magnetic, electric, and microwave fields. In case of strong qubit-environment couplings, pumped point defect qubit systems serve as efficient spin polarization sources that can be utilized in hyperpolarization applications\citep{Broadway2018,Wunderlich_2018} either for enhancing the sensitivity of magnetic resonance experiments or for cooling environmental spins to reduce local magnetic field fluctuations.
    
Deeper understanding and numerical description of decoherence, spin relaxation, and polarization transfer over a wide range of environmental conditions are essential for advanced future applications. Lindblad master equation that describes Markovian decay processes is frequently applied when dynamical properties are considered. On the other hand, this approach relies on experimental decay rates and neglects the complexity of environmental interactions that may cause loss of quantitative accuracy and predictive power. To overcome these limitations numerous theoretical studies have been recently reported in this subject.

Several powerful theoretical tools have been developed to describe decoherence processes. For example, quantum cluster expansion\citep{Witzel2005,Witzel:PRB2006}, linked cluster expansion\citep{ShamLinkedCluster2007}, nuclear pairwise model\citep{Liu_2007}, disjoint cluster model\citep{MazePRB2008Decoh}, semi-classical magnetic field approximation\citep{Taylor:NatPhys2008,Hanson:Science2008}, ring diagram approximation\citep{Sarma2009}, analytic approaches\citep{Maze_2012,HollenbergDecoh2014,BalianPRB2014}, spin-coherent P-representation method\citep{Susumu2013} and cluster-correlation expansion (CCE)\citep{YangPRB2008CCE,YangPRB2009CCEEnsamble,RenBaoLiuNVDecoh2012,YangPRB2014,SeoNatureComm2016}  
have been utilized to calculate T$_2$ and T$_2^{\star}$ times of point defect qubit systems. Temperature dependence of spin-phonon-coupling induced spin relaxation of NV center was recently studied by analytic\citep{DOHERTY20131,Doherty2012,Norambuena2018} and \emph{ab initio}\citep{GuglerT1T2018} approaches. Theoretical studies on spin bath induced spin relaxation processes have focused on strong environmental coupling regions where dynamical nuclear polarization can be achieved.\citep{,IvadyDNP2015,IvadyPRL2016,Wunderlich2017,Broadway2018,ANISHCHIK201967}  Much less attention has been paid, however, to the calculation of spin bath assisted relaxation processes and related decay time T$_1$ of point defect qubits at general control parameter settings where spin flip-flops are suppressed to a large degree. In a very recent study, CCE method was generalized to describe spin flip-flops of a NV center interacting with a bath of $^{13}$C nuclear spins.\citep{Yang2019} Time-dependent mean field algorithm \citep{AlHassaniehPRL2006} applied successfully to quantum dot systems\citep{SinitsynPRL2012} is a promising alternative approach.

The rest of the article is organized as follows. Section~\ref{sec:method} describes the formulation of the theoretical approach and details of the implementation. Section~\ref{sec:dyn} discusses time evolution of an exemplary spin system at different level of approximation.  Section~\ref{sec:sum_res} provides numerical results  on the spin relaxation of NV center in diamond. Finally, section~\ref{sec:disc} summarizes  and concludes our findings.

\section{Methodology}
 \label{sec:method}

In this section we discuss cluster approximation of a many particle system in the framework of an extended Lindblad formalism to simulate spin relaxation processes. Hereinafter, we use the following terminology. We regard subsystems of a closed or open system as spins. Spins are either elementary building blocks of the system or complex, many-level systems. In the latter case spins can be defined based on the difference of internal and external coupling strength. We assume that inter-spin couplings are weaker than intra-spin couplings. Furthermore, we name processes that change the diagonal elements of spins' reduced density matrices as spin flip-flop processes.

\subsection{First order cluster approximation} \label{sec:fo_cluster_approx}

\begin{figure}[h!]
	\includegraphics[width=0.50\columnwidth]{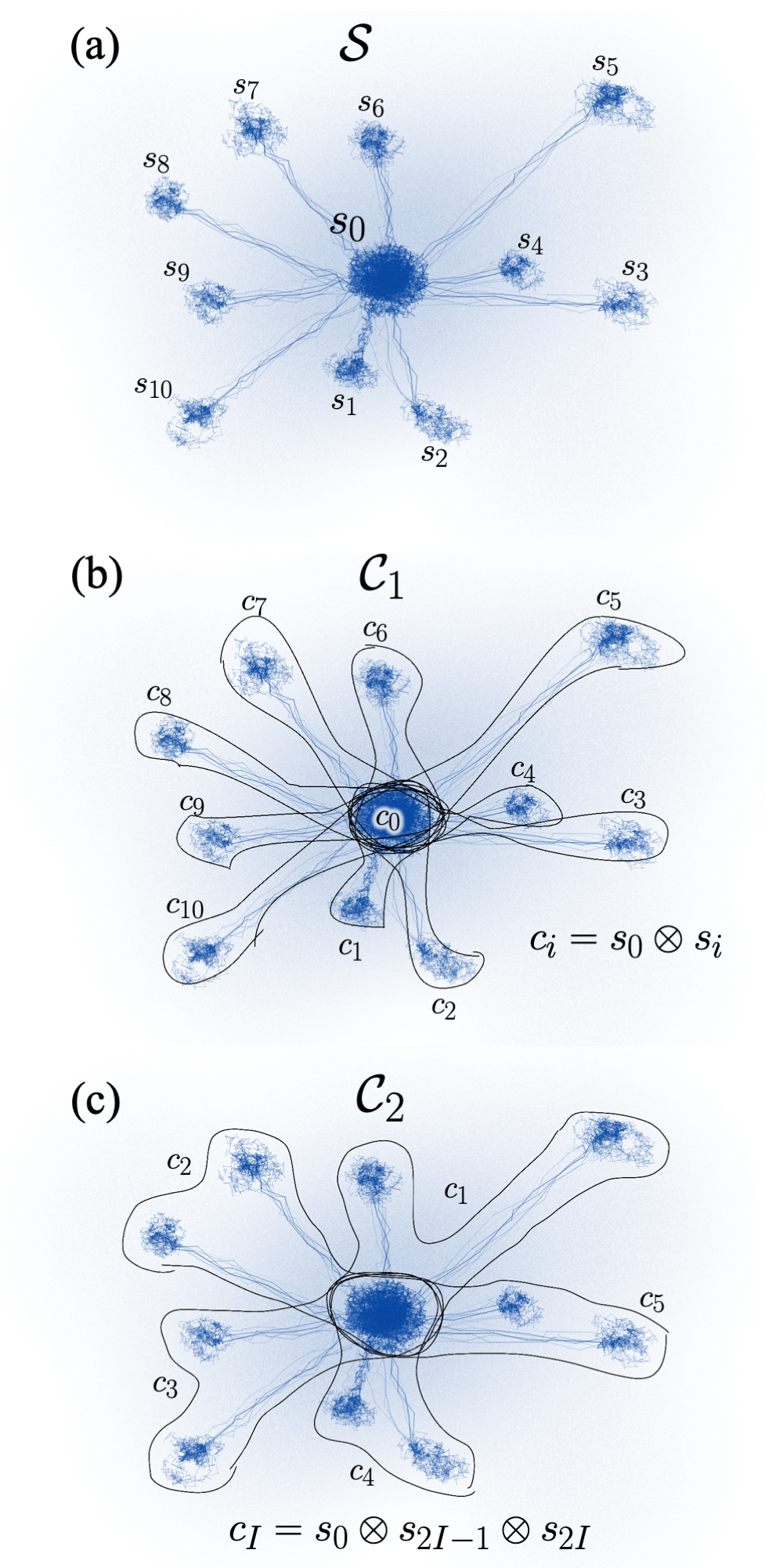}
	\caption{ Cluster approximations of a many-spin system $\mathcal{S}$. (a) $\mathcal{S}$ consists of a central spin $s_0$ and number of $n$ coupled spins $s_i$ that couple only to the central spin $s_0$.   (b) First order cluster approximation of $\mathcal{S}$ that comprises $n+1$ cluster systems $c_0$ and $c_i$.  $c_0$ includes the central spin $s_0$ only, while $c_i$ for $i \neq 0$ includes a pair of spins, $s_0$ and $s_i$. (c) Second order cluster approximation of $\mathcal{S}$ that comprises $n/2 + 1$ cluster systems $c_I$, where each cluster system contains $s_0$ and two coupled spins $s_I$ where $1 \leq I \leq n/2$.  $c_0$ includes solely the central spin $s_0$. }
	\label{fig:c0}  
\end{figure}

Let us consider an open system $\mathcal{S}$ that consists of a central spin $s_0$ and a number of environmental or bath spins $s_i$, where $i = \left\lbrace 1,...,n \right\rbrace$. Furthermore, let us denote the dimension of the Hilbert space of the central and environmental spins by $d_0$ and $d_i$, respectively. First, we assume that $s_i$ couples only to the central spin, see Fig.~\ref{fig:c0}(a).

The master equation of the open system $\mathcal{S}$ can be written as
\begin{equation} \label{eq:MasterS}
\dot{\varrho}_{\mathcal{S}} = - \frac{i}{\hbar} \left[ H_0, \varrho_{\mathcal{S}} \right] + \mathcal{E} \! \left( \varrho_{\mathcal{S}} \right) \text{,}
\end{equation}
where the Hamiltonian $H_0$ can be written as
\begin{equation}
H_0 = h_{\text{0} } +  \sum_{i=1}^n \left(  h_{ i } + h_{0 i } \right)  \text{,}
\end{equation}
where $ h_{0} $ is the Hamiltonian of the central spin, $ h_{i} $ is the Hamiltonian of the coupled spin $s_i$, and  $ h_{0i } $ describes the coupling of the central spin and the bath spin $s_i$. The last term on the right hand side of Eq.~(\ref{eq:MasterS}) accounts for environmental effects, such as temperature dependent effects and spin relaxation due to spins that are not included in $\mathcal{S}$, through the Lindbladian $\mathcal{E}$. 
The size of the  problem, i.e the dimension of the Hilbert space, increases exponentially with $n$, which makes an exact solution unfeasible for large $n$.

To model the dynamics of $\mathcal{S}$ we divide it into a cluster $\mathcal{C}_{\mathcal{N}}$ of overlapping cluster systems, where $\mathcal{N}$ is the order of the cluster approximation. In first order cluster approximation cluster $\mathcal{C}_1$ consists of $n+1$ cluster systems $c_0$ and $c_i$, where $i \in \left\lbrace 1,...,n \right\rbrace$. Expect for $c_0$, which includes only $s_0$, all other cluster systems include the central spin and one coupled spin $s_i$, see Fig.~\ref{fig:c0}(b) for illustration.  Hamiltonians of the cluster systems can be written as, 
\begin{eqnarray} \label{eq:h_c0}
&  h_{c_0} = h_0 \text{,} \\
\label{eq:h_ci}
&  h_{c_i} = h_0 + h_i + h_{0i} \text{.}
\end{eqnarray} 

We may rationalize the above clustering by considering each $c_i$ cluster system as an implement to measure spin flip-flops induced by the coupled spin $s_i$. Cluster system $c_0$ serves as a reference system where the central spin evolves freely without interacting with other spins.

Master equations of the cluster systems can be written as
\begin{equation}\label{eq:ro_ci_master}
\dot{\varrho}_{c_0} = - \frac{i}{\hbar} \left[ h_{c_0}, \varrho_{c_0} \right]  + \mathcal{E}_{c_0} \! \left( \varrho_{c_0} \right)  \text{,} \ \ \text{ and   } \ \  \dot{\varrho}_{c_i} = - \frac{i}{\hbar} \left[ h_{c_i}, \varrho_{c_i} \right] + \mathcal{E}_{c_i} \! \left( \varrho_{c_i} \right) \text{,}
\end{equation}
where the dimensions of the density matrices are given by $\text{dim}\! \left( \varrho_{c_i} \right) =  d_0 d_i $. $\mathcal{E}_{c_0}$ and $\mathcal{E}_{c_i}$ describe environmental effects not induced by the spin bath of $\mathcal{S}$. The density matrix of  the coupled spin $s_{i}$ can be determined by tracing over $s_0$ in $c_i$,
\begin{equation}
\varrho_{s_i} = \text{Tr}_0 \! \left( \varrho_{c_i} \right) \text{.}
\end{equation}
As the central spin is included in all cluster systems, there are altogether $n+1$ definitions for the reduced density matrix of the central spin, i.e.
\begin{equation}
\varrho_{s_0}^{c_0} =  \varrho_{c_0}   \ \ \text{ and   } \ \   \varrho_{s_0}^{c_i} = \text{Tr}_{i} \! \left( \varrho_{c_i} \right) \text{.}
\end{equation}

In the following subsections, we introduce couplings between the cluster systems to approximate the dynamics of the many spin system $\mathcal{S}$. 
First, we extend Eq.~(\ref{eq:h_c0}) and Eq.~(\ref{eq:h_ci}) to account for an effective intra-spin bath field that time dependently shifts energy eigenvalues and preserves diagonal elements of the reduced density matrix of the central spin.  Second, we extend Eqs.~(\ref{eq:ro_ci_master}) by additional time dependent Lindbladian terms to account for interactions that induce spin flip-flops and cause variation of the diagonal elements of the central spin's density matrix.

\subsubsection{Mean intra-spin bath field} \label{sec:ext_mif}

The interaction Hamiltonian $h_{0i}$ may include terms that do not induce spin flip-flops of the central spin but rather shift the energy levels. As such interactions alter the energy level structure of the system, they may  affect the dynamics of the central spin too. According to Eq.~(\ref{eq:h_ci}) and Eq.~(\ref{eq:ro_ci_master}), cluster system $c_i$ describes energy shifts solely due to spin-bath spin $s_i$, as the Hamiltonian $h_{c_i}$ does not depend on other spin-bath spin degrees of freedom. Energy shifts due to other spins, however, may be taken into account by introducing an effective field acting on the central spin. This field is of course different in all cluster systems.

In order to account for the effective field of environmental spins included in other cluster systems, we extend $h_0$ as
\begin{equation} \label{eq:h0_ext}
h_0^{c_0} = h_0 + \beta_0 \ \ \  \text{ and } \ \ \  h_0^{c_i} = h_0 + \beta_i \text{,}
\end{equation} 
where $\beta_0$ and $\beta_i$ describe effective fields acting on $s_0$ in cluster system $c_0$ and $c_i$, respectively. To define $\beta_i$, we first calculate the internal field $\alpha_i$ in each cluster system $c_i$ obtained from the polarization of the environmental spin $s_i$ through a semi-classical formula
\begin{equation} \label{eq:alpha_i}
\alpha_i =\text{Tr}_i \! \left( h_{0i} \circ \left( I_{d_0} \otimes \text{Tr}_0 \varrho_{c_i}  \right) \right) \text{,}
\end{equation}
where $I_{d_0}$ is the identity matrix of $d_0$ dimension and $\left( A \circ B \right)_{mn} = A_{mn}B_{mn} \delta_{mn} $, where $\delta_{mn} $  is the Kronecker delta. To elucidate this definition, let us assume that the interaction Hamiltonian $h_{0i}$ contains a single term $ \gamma S_z^{s_0} S_z^{s_i} $, where $\gamma$ is the coupling strength and $S_z^{s_0} $ and $S_z^{s_i} $ are spin $z$ operators. For this Hamiltonian $\alpha_i$ is equal to  $ \gamma  \left\langle S_z^{s_i} \right\rangle  S_z^{s_0} $, where $\left\langle S_z^{s_i} \right\rangle $ is the expectation value of $ S_z^{s_i}$. 

From $\alpha_i$ we can define the effective field of environmental spins included in other cluster systems as
\begin{equation} \label{eq:beta_0}
\beta_0 = \sum_{i=1 }^n  \alpha_i
\end{equation}
and
\begin{equation} \label{eq:beta_i}
\beta_i = \sum_{j=1, j \neq i}^n \left( \alpha_j \otimes I_{d_i} \right) \text{,}
\end{equation}
where $I_{d_i} $ is the identity matrix of $d_i $ dimension. The extended Hamiltonians of the cluster systems can be written as, 
\begin{eqnarray} \label{eq:h_c0_ext}
&  \widetilde{h}_{c_0} = h_0^{c_0} \text{,} \\
\label{eq:h_ci_ext}
&  \widetilde{h}_{c_i} = h_0^{c_i} + h_i + h_{0i} \text{.}
\end{eqnarray} 

We note that the effective internal field defined by Eq.~(\ref{eq:beta_0}) and Eq.~(\ref{eq:beta_i}) act solely on the central spin. Note furthermore that the total effective field in each cluster systems is equal to $\beta_0$ as $\text{Tr}_i \! \left( \beta_i + \alpha_i \right) = \beta_0$. When a nuclear spin bath is considered, the effective field $\beta_0$ of the polarized nuclear spin bath may be referred to as the Overhauser field. Finally, note that the internal effective field can be utilized to account for dephasing effects in a semi classical approximation. Study of such processes is outside the scope of the present article. 

\subsubsection{Extended Lindbladian} \label{sec:ext_lindb}

Faithful description of spin flip-flops of the central spin due to the interaction with the spin bath requires additional extension. Without coupling between the cluster systems, the central spin $s_0$ in a cluster system $c_i$ undergoes environmental spin induced flip-flops that are solely driven by environmental spin $s_i$. In order to simulate the dynamics of the many spin system $\mathcal{S}$, we require through a non-unitary coupling between the cluster systems that the central spin in all cluster systems undergoes spin flip-flops induced by all the environmental spins. This effectively ensures that  the diagonal elements of the reduced density matrix of the central spin are identical in all cluster systems.To this end we introduce an extended, time dependent Lindbladian. 

First of all, we extend the master equations of cluster systems $c_0$ and $c_i$ by adding time dependent Lindbladian terms $\mathcal{L}_{c_0}$ and $\mathcal{L}_{c_i}$, as
\begin{equation} \label{eq:ext_linbd_0}
\dot{\varrho}_{c_0} = - \frac{i}{\hbar} \left[ \widetilde{h}_{c_0}, \varrho_{c_0} \right] + \mathcal{L}_{c_0} \! \left( \varrho_{c_0} \right) + \mathcal{E}_{c_0} \! \left( \varrho_{c_0} \right) 
\end{equation}
and
\begin{equation} \label{eq:ext_linbd_i}
\dot{\varrho}_{c_i} = - \frac{i}{\hbar} \left[ \widetilde{h}_{c_i}, \varrho_{c_i} \right] + \mathcal{L}_{c_i} \! \left( \varrho_{c_i} \right) + \mathcal{E}_{c_i} \! \left( \varrho_{c_i} \right) \text{,}
\end{equation}
where the Lindbladians are defined in the form of
\begin{equation}\label{eq:lindbladian_0}
 \mathcal{L}_{c_0} \!\! \left( \left\lbrace \dot{b}_{0l} \right\rbrace ,  \left\lbrace C_{0l} \right\rbrace; \varrho_{c_0} \right)   =\sum_{l} \frac{\dot{b}_{0l}}{\text{Tr} \! \left( C_{0l}^\dagger C_{0l}  \varrho_{c_0} \right)} \left( C_{0l}  \varrho_{c_0} C_{0l}^\dagger -  \frac{1}{2} \left( \varrho_{c_0} C_{0l}^\dagger C_{0l} + C_{0l}^\dagger C_{0l}  \varrho_{c_0} \right) \right) 
\end{equation}
and 
\begin{equation}\label{eq:lindbladian_i}
 \mathcal{L}_{c_i} \!\! \left( \left\lbrace \dot{b}_{il} \right\rbrace ,  \left\lbrace C_{il} \right\rbrace , \varrho_{c_i} \right)   =\sum_{l} \frac{\dot{b}_{il}}{\text{Tr} \! \left( C_{il}^\dagger C_{il}  \varrho_{c_i} \right)} \left( C_{il}  \varrho_{c_i} C_{il}^\dagger -  \frac{1}{2} \left( \varrho_{c_i} C_{il}^\dagger C_{il} + C_{il}^\dagger C_{il}  \varrho_{c_i} \right) \right)  \text{,}
\end{equation}
where $\dot{b}_{0l}$ and $\dot{b}_{il} \ge 0$ are time dependent rates and $C_{0l}$ and  $C_{il}$ are Lindblad operators. 
We consider $C_{0l}$  and  $C_{il}$ operators that describe solely spin flip and flop transitions of the central spin. Therefore, $C_{0l}$ and $C_{il}$ operators can be written as 
\begin{equation}\label{eq:def_Cil}
C_{0l} = C_l  \ \ \text{ and }  \  \ C_{il} = C_l \otimes I_i \text{,}
\end{equation}
where $C_l$ Lindblad operators of $d_0$ dimension are identical for all cluster systems. Altogether $\ d_0 \left( d_0 -1\right)$ number of independent $C_{l}$ operators can be defined. We define these operators as
\begin{equation} \label{eq:def_Cl}
C_l  = \left| m \right\rangle \! \left\langle n \right| 
\end{equation} 
where $ \left| m \right\rangle $ and $\left| n \right\rangle$ are states of an orto-normal basis that spans the Hilbert space of $s_0$ and $m \neq n$. Hereinafter, we use $l$ index as a shorthand notation of $mn$ indices. Note that $\left\lbrace C_l \right\rbrace$ includes operators that drive spin flip-flops both forward and backward, i.e.\  $C_k$ and  $C_k^{\dagger} \in \left\lbrace C_l \right\rbrace$. This condition is required by the irreversible effect of the  extended Lindbladians in Eq.~(\ref{eq:lindbladian_0}) and (\ref{eq:lindbladian_i}) and the positivity of $\dot{b}_{0l}$ and $\dot{b}_{il}$ rates. Furthermore, we note that Eq.~(\ref{eq:lindbladian_0}) and (\ref{eq:lindbladian_i}) require that
\begin{equation}
\text{Tr} \! \left( C_{il}^\dagger C_{il}  \varrho_{c_i} \right) = \text{Tr} \! \left(  \left| n \right\rangle \! \left\langle n \right|  \varrho^{c_i}_{s_0} \right) = \left( \varrho^{c_i}_{s_0} \right)_{nn} \neq 0 \text{,}
\end{equation}
i.e.\  $\left| m \right\rangle \! \left\langle n \right| $ spin flip-flop processes are only possible when the population in the initial $\left| n \right\rangle $ state is non-zero. To explicitly handle the exception when $\text{Tr} \! \left( C_{il}^\dagger C_{il}  \varrho_{c_i} \right) = 0$, we define $ {\dot{b}_{il}}  {\text{Tr}^{-1} \! \left( C_{il}^\dagger C_{il}  \varrho_{c_i} \right)} = 0$. 

Furthermore, we draw attention to a specific property of the definitions
\begin{equation} \label{eq:prop1}
dt  \mathcal{L}_{c_0} \!\! \left( \left\lbrace \dot{b}_{0l} \right\rbrace ,  \left\lbrace C_{0l} \right\rbrace ; \varrho_{c_0}  \right)_{kk} = \sum_{n \neq k}  dt \dot{b}_{0(kn)} -  \sum_{m \neq k}  dt \dot{b}_{0(mk)}  
\end{equation}
and 
\begin{equation} \label{eq:prop2}
dt  \mathcal{L}_{c_i} \!\! \left( \left\lbrace \dot{b}_{il} \right\rbrace ,  \left\lbrace C_{il} \right\rbrace ; \varrho_{c_i}  \right)_{kk} = \sum_{n \neq k}  dt \dot{b}_{i(kn)} -  \sum_{m \neq k}  dt \dot{b}_{i(mk)}   
\end{equation}
where we explicitly use $ l = (mn)$ indices and $dt $ is an infinitesimal time period. The right hand side of Eq.~(\ref{eq:prop1}) and  Eq.~(\ref{eq:prop2}) describe how the diagonal elements of the density matrix $\varrho_{c_0} $ and $\varrho_{c_i} $ change due to $\mathcal{L}_{c_0} $ and $\mathcal{L}_{c_i} $ over $dt$ propagation, respectively. Note that the variation of the diagonal elements is irrespective of the density matrix and determined solely by time dependent rates $\dot{b}_{0l}$ and $\dot{b}_{il}$. 
 
We utilize $ \mathcal{L}_{c_0}$ and $ \mathcal{L}_{c_i}$ Lindbladians to carry out such spin flip-flops of the central spin in $c_0$ and $c_i$ that happen in cluster system $c_j$ due to coupling to $s_j$ for $j \neq i$. This effectively makes the diagonal elements of the reduced density matrix of the central spin to be identical in all cluster systems during the time evolution, i.e.
\begin{equation}\label{eq:rho_so_eq}
 \text{diag }\! \varrho_{s_0}^{c_0}  = \text{diag }\! \varrho_{s_0}  \text{,}  \ \ \text{ and } \ \ \text{diag }\! \varrho_{s_0}^{c_i} = \text{diag }\! \varrho_{s_0} 
\end{equation}
for any $i$ at any time $t$, where $\text{diag }\! \varrho$ is the vector of diagonal elements of $\varrho$. We utilize time dependent rates $\dot{b}_{0l}$ and $\dot{b}_{il}$ in Eqs.~(\ref{eq:ext_linbd_0})-(\ref{eq:ext_linbd_i}) to achieve this goal. 

Before defining $\dot{b}_{0l}$ and $\dot{b}_{il}$, we need to quantify internal flip-flop rates in each cluster systems.  To do so, we define $\dot{a}_{0l}$ and $\dot{a}_{il}$ positive rates in such a way that the following equality are satisfied,
\begin{eqnarray} \label{eq:a0l_def}
 \text{diag} \! \left(  e^{- \frac{i}{\hbar} \left( h_{c_0} . - . h_{c_0}  \right) t }\varrho_{c_0} \! \left(  t = 0 \right)   \right) =  \text{diag }\! \varrho_{s_0}^{c_0}  \! \left( t \right) =  \nonumber \\
  \text{diag } \! \! \left(  \varrho_{c_0} \! \left(  t = 0 \right) +  \int_0^t \mathcal{L}_{c_0} \!\! \left( \left\lbrace \dot{a}_{0l} \! \left( \tau \right) \right\rbrace \text{,}  \left\lbrace C_{0l} \right\rbrace ;  \varrho_{c_0} \! \left(  \tau \right) \right)   d\tau   \right)   \text{,}
\end{eqnarray}
and 
\begin{eqnarray}  \label{eq:ail_def}
 \text{diag Tr}_i \! \left(  e^{- \frac{i}{\hbar} \left( h_{c_i} . - . h_{c_i}  \right) t }\varrho_{c_i} \! \left(  t = 0 \right)  \right) =   \text{diag }\! \varrho_{s_0}^{c_i} \! \left( t \right) =  \nonumber \\
  \text{diag } \text{Tr}_i \! \! \left(   \varrho_{c_i} \! \left(  t = 0 \right) +  \int_0^t   \mathcal{L}_{c_i} \!\! \left( \left\lbrace \dot{a}_{il} \! \left( \tau \right) \right\rbrace \text{,}  \left\lbrace C_{il} \right\rbrace ;  \varrho_{c_i} \! \left( \tau \right) \right)   d\tau \right)  \text{.}
\end{eqnarray}
Note that the parentheses on the left hand side of Eq.~(\ref{eq:a0l_def}) and Eq.~(\ref{eq:ail_def}) contain the general solution of Eqs.~(\ref{eq:ro_ci_master}), while the parentheses on the right hand side of Eq.~(\ref{eq:a0l_def}) and Eq.~(\ref{eq:ail_def}) contain the general solution of $\dot{\varrho}_{c_0} = \mathcal{L}_{c_0} \!\! \left( \left\lbrace \dot{a}_{0l}  \right\rbrace \text{,}  \left\lbrace C_{0l} \right\rbrace ; \varrho_{c_0}\right)  $ and $\dot{\varrho}_{c_i} = \mathcal{L}_{c_i} \!\! \left( \left\lbrace \dot{a}_{il}  \right\rbrace \text{,}  \left\lbrace C_{il} \right\rbrace ; \varrho_{c_i} \right)  $, respectively. The above equality ensure that the time dependence of the diagonal elements of the density matrix of $s_0$ can be obtained by evaluating either the right hand side or the left hand side of Eq.~(\ref{eq:a0l_def}) and Eq.~(\ref{eq:ail_def}). Rates $\dot{a}_{0l} $ and $\dot{a}_{il}$ that fulfill Eq.~(\ref{eq:a0l_def}) and Eq.~(\ref{eq:ail_def}) thus determine flip-flops rates of the central spin due to the corresponding Hamiltonian $h_{c_0}$ and $h_{c_i}$ of the cluster systems. For practical reasons, calculation of $\dot{a}_{0l} $ and $\dot{a}_{il}$ rates may include additional simplifications and approximations, see section~\ref{sec:imp}.

To measure differences of the spin flip-flop rates between cluster systems $c_0$ and $c_i$ during the time evolution, we calculate
\begin{equation} \label{eq:dail}
 \Delta \dot{a}_{il} = \dot{a}_{il} - \dot{a}_{0l}  \  \text{ for }  \  \dot{a}_{il} > \dot{a}_{0l} \ \ \ \text{ and } \ \ \ 
 \Delta \dot{a}_{il} = 0 \  \text{ for }  \  \dot{a}_{il} < \dot{a}_{0l} \text{.} 
\end{equation}
The role of cluster system $c_0$ that includes only the central spin is apparent from Eq.~(\ref{eq:dail}). As $s_0$ in $c_0$ interacts with no environmental spin directly, $a_{0l}$ measure flip-flop rates intrinsic to the central spin.  Thus $ \Delta \dot{a}_{il}$ quantifies spin flip-flop rates of the central spin induced solely by environmental spin $s_i$ in cluster system $c_i$. 

Finally, let us define the time dependent rates $\dot{b}_{0l}$ and $\dot{b}_{il} $ entering  Eq.~(\ref{eq:lindbladian_0}) and Eq.~(\ref{eq:lindbladian_i}) as
\begin{equation} \label{eq:b0}
\dot{b}_{0l} = \sum_{i=1}^{n} \Delta \dot{a}_{il}   
\end{equation}
and 
\begin{equation} \label{eq:bi}
\dot{b}_{il} = \sum_{j=1, j \neq i}^{n} \Delta \dot{a}_{jl}   \text{,}
\end{equation} 
respectively. $\dot{b}_{0l} $ determine spin flip-flop rates of the central spin induced by all the environmental spins, while $\dot{b}_{il} $ determine spin flip-flop rates of the central spin induced by  environmental spins other than $s_i$.

The central assumption of the proposed method is that the self consistent solution of Eq.~(\ref{eq:ext_linbd_0}) and Eq.~(\ref{eq:ext_linbd_i}) with the time dependent Lindbladian given in Eq.~(\ref{eq:lindbladian_0}) and Eq.~(\ref{eq:lindbladian_i}) coupled through the time dependent rates defined in Eq.~(\ref{eq:b0}) and Eq.~(\ref{eq:bi}) approximately describes spin flip-flop processes of the many spin system $\mathcal{S}$. A cornerstone of the approximation in Eq.~(\ref{eq:b0}) and Eq.~(\ref{eq:bi}) is the additivity of the time dependent rates $\Delta a_{0l}$ and $\Delta a_{il}$. In Appendix~\ref{app:sum_deriv}, we demonstrate that additivity is a good approximation for a non-entangled or partially entangled central spin system over an infinitesimal $dt$ time evolution. Note that the first order cluster approximation, that neglects entanglement within the spin bath, and self-consistent solution of the equations ensure that the additivity holds at any time $t$ during the time evolution of $\mathcal{C}_1$. 

It is clear from the above discussion that the main approximation in the description of the central spin-spin bath coupling is the assumption of a non-entangled spin bath. This may be a good approximation when the coherence time of the spin bath specimens is shorter than the inverse coupling strength between the central spin and the spin bath specimens, i.e.\
\begin{equation}
T_2^{i} << \frac{1}{ \left| h_{0i} \right| } \text{.}
\end{equation}
As we will see in the numerical calculations, this approximation is either satisfied or not satisfied depending on the type of environmental spins. Note, however, that in the latter case the approximation can be systematically improved by including spin bath interactions and inter spin bath correlations in higher order cluster approximations, see section~\ref{sec:ho_cluster_approx}.   

It is apparent from the equations that the method does not require additional approximation of the Hamiltonian beyond the central spin approximation. Furthermore, restrictions due to central spin approximation can be remedied by using higher order cluster approximations, see section~\ref{sec:ho_cluster_approx}. Note however that the approximation depends somewhat on the choice of the basis states used to span the Hilbert space of $s_0$. In addition, the formalism ensures that conservable quantities are conserved in the cluster model. Let us assume that the $h_0$ and $h_i$ are defined so that extensive quantity $E$ is preserved in all cluster systems $c_0$ and $c_i$. Properties shown in  Eqs.~(\ref{eq:prop1})-(\ref{eq:prop2}) and the summation in Eq.~(\ref{eq:b0}) and Eq.~(\ref{eq:bi}) make $E$ conserved in the cluster model $\mathcal{C}_1$ too. Finally, we note that phase information of the central spin can be lost due to the non-unitary Lindbladian drive utilized in the method. Due to this and the conservation property, $\mathcal{C}_1$ may be considered as a partially open model of the  many spin system $\mathcal{S}$.

\subsection{Higher order cluster approximations} \label{sec:ho_cluster_approx}

In higher order cluster approximation $\mathcal{C}_N$, where $N$ is the order parameter, cluster systems $c_I$  for $I \in \left\lbrace 1, ... , n/N \right\rbrace$ contain a number of $N$ environmental spins. Cluster system $c_0$ contains the central spin only. The central spin is included in all other cluster systems and each of the  environmental spins is included only in a single cluster system as illustrated in Fig.~\ref{fig:c0}(c). While first order cluster approximation is unique in all cases, higher order cluster approximations can be defined differently depending on how the environmental spin are clustered. For simplicity, here we assume that clustering is based on the indices of the environmental spins.

The Hamiltonian of higher order cluster system $c_I$ can be defined as
\begin{equation} \label{eq:h_cI}
 h_{c_I} = h_0 + \sum_{i = 1+ N \left( I - 1 \right) }^{NI} \left( h_i + h_{0i} \right) + \sum_{i = 1+ N \left( I - 1 \right), \   j = i+1 }^{NI} h_{ij} \text{,}
\end{equation}  
where $h_{ij}$ describes interactions of environmental spins. The Lindblad equation of the density matrix $\varrho_{c_I}$ of cluster system $c_I$ can be written as
\begin{equation}\label{eq:ro_cI_master}
 \dot{\varrho}_{c_I} = - \frac{i}{\hbar} \left[ h_{c_I}, \varrho_{c_I} \right] + \mathcal{E}_{c_I} \! \left( \varrho_{c_I} \right) \text{.}
\end{equation}

We note that definitions for the reference system $c_0$ are the same in every order of the cluster approximation. Furthermore, the definitions and approximations introduced in section~\ref{sec:ext_mif} and \ref{sec:ext_lindb} are irrespective of the order of the cluster approximation. In general the corresponding definitions can be obtained by substituting index $i$ with index $I$.  

Note that the approximations introduced in section~\ref{sec:ext_mif} and \ref{sec:ext_lindb} are systematically improvable by increasing the clustering order. Larger cluster systems describe coupling and entanglement that are completely neglected in first order cluster approximation. Ultimately, for $N = n $ we return to the exact case.

\subsection{Numerical implementation}
\label{sec:imp}

\begin{figure}[h!]
	\includegraphics[width=0.90\columnwidth]{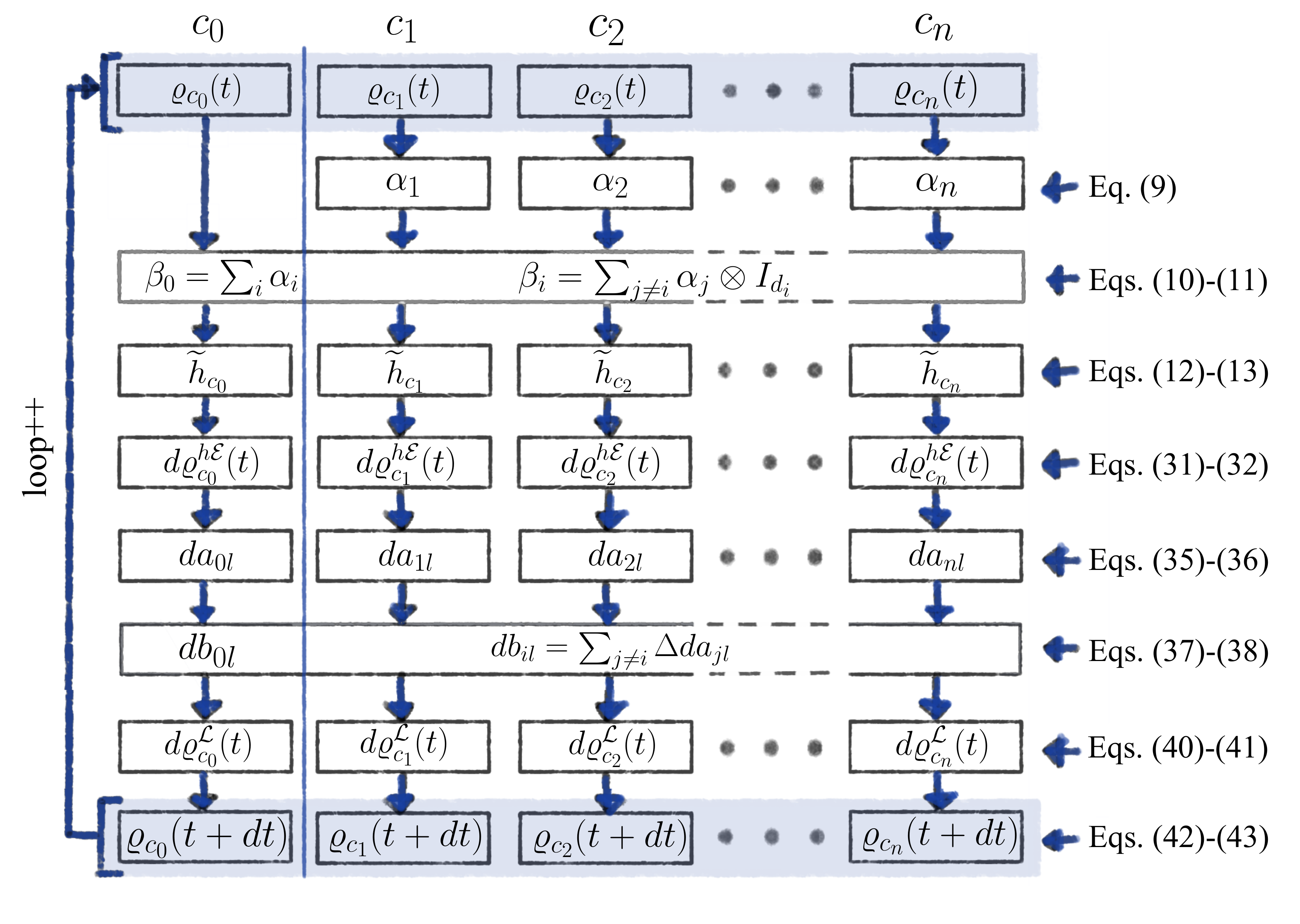}
	\caption{ Time propagation cycle. Each rectangle indicates a computational task. Tasks are calculated either parallel for all the cluster systems (individual rectangles) or by using common operations (shared rectangles).  }
	\label{fig:loop}  
\end{figure}

While analytic solution of the extended coupled Lindblad equations in cluster approximation  is hard even for simple systems, numerical propagation of the model is straightforward and can be efficiently implemented for  parallel computing.   In Fig.~\ref{fig:loop}, we schematically summarize the most important computational steps for sequential propagation of the cluster model. Certain steps can be calculated in parallel while others require the calculation of common quantities of the systems, see Fig.~(\ref{fig:loop}). Next, we go through the time propagating cycle step by step. 

(i) Let us assume that at time $t$ the density matrices of the cluster systems $\varrho_{c_0} \! \left( t \right)$ and $\varrho_{c_i} \! \left( t \right)$ are given. (ii) The internal effective field $\alpha_{i}$ is calculated  through Eq.~(\ref{eq:alpha_i}) in all cluster systems $c_i$ that include an environmental spin. (iii) From $\alpha_{i}$, the effective field of the spin bath $\beta_0$ is calculated for cluster system $c_0$ and effective fields of environmental spins $j \neq i $ are calculated for cluster systems $c_i$ through Eq.~(\ref{eq:beta_0}) and Eq.~(\ref{eq:beta_i}), respectively. (iv) Hamiltonians of the cluster systems are calculated from Eq.~(\ref{eq:h0_ext}) and Eq.~(\ref{eq:h_c0_ext}) and Eq.~(\ref{eq:h_ci_ext}). 

After the cluster dependent part of the Hamiltonians is determined, in step (v) the cluster systems are propagated according to their independent master equations, given in Eq.~(\ref{eq:ro_ci_master}), over a short period of time $dt$  in order to obtain the variation of the density matrix $d \varrho^{h\mathcal{E}}_{c_0}$ and $d \varrho^{h\mathcal{E}}_{c_i}$ caused by Hamiltonian $\widetilde{h}_{c_i}$ and Lindbladian $\mathcal{E}_{c_i}$, i.e.
\begin{equation}
 d \varrho^{h\mathcal{E}}_{c_0} \! \left( t \right)  = - \frac{idt}{\hbar} \left[ \widetilde{h}_{c_0},  \varrho_{c_0} \! \left( t \right) \right] + dt \mathcal{E}_{c_0} \! \left(   \varrho_{c_0} \! \left( t \right) \right)  
\end{equation}
and 
\begin{equation}
 d \varrho^{h\mathcal{E}}_{c_i} \! \left( t \right)  =  - \frac{idt}{\hbar} \left[ \widetilde{h}_{c_i}, \varrho_{c_i} \! \left( t \right) \right] + dt \mathcal{E}_{c_i} \! \left(   \varrho_{c_i} \! \left( t \right) \right)    \text{.}
\end{equation}
To eliminate errors up to $O \! \left( dt^5 \right)$ we utilize Runge-Kutta method in this step.

In step (vi) of the propagation cycle, we quantify in each cluster system the spin flip-flops occurred during the short propagation calculated in the previous step. To do so, we restrict Eq.~(\ref{eq:a0l_def}) and Eq.~(\ref{eq:ail_def}) to infinitesimal time evolution and obtain 
\begin{equation} \label{eq:da0l_def}
 \text{diag }  d\varrho_{c_0}^{h\mathcal{E}} \! \left( t \right)  = \text{diag }   \mathcal{L}_{c_0} \!\! \left( \left\lbrace da_{0l} \right\rbrace \text{,}  \left\lbrace C_{0l} \right\rbrace ; \varrho_{c_0}  \! \left( t \right)  \right)  
\end{equation}
and
\begin{equation} \label{eq:dail_def}
\text{diag } \text{Tr}_i  d\varrho_{c_i}^{h\mathcal{E}} \! \left( t \right)   = \text{diag } \text{Tr}_i   \mathcal{L}_{c_i} \!\! \left( \left\lbrace da_{il}  \right\rbrace \text{,}  \left\lbrace C_{il} \right\rbrace ;  \varrho_{c_i}  \! \left( t \right) \right)   \text{,}
\end{equation}
where $da_{0l} = \dot{a}_{0l} dt$ and  $da_{il} = \dot{a}_{il} dt$. 
    
Before discussing the next step of the cycle, we discuss through a few examples how to obtain $ da_{0l}$ and $ da_{il}$ in practice. $da_{0l} $ and $da_{il} $ describe infinitesimal population transitions of the diagonal elements of $ \varrho_{c_0} $ and $ \varrho_{c_i} $, respectively.  Based on Eqs.~(\ref{eq:prop1})-(\ref{eq:prop2}), we can rewrite Eq.~(\ref{eq:da0l_def}) as
\begin{equation}\label{eq:da0l_eq2}
\left( d\varrho_{c_0}^{h\mathcal{E}} \right)_{jj} = \sum_{n \neq j} da_{0(jn)} - \sum_{m \neq j} da_{0(mj)}  \text{,}
\end{equation}
and  Eq.~(\ref{eq:dail_def}) as
\begin{equation}\label{eq:dail_eq2}
\left( \text{Tr}_i d\varrho_{c_i}^{h\mathcal{E}} \right)_{jj} = \sum_{n \neq j} da_{i(jn)} - \sum_{m \neq j} da_{i(mj)}  \text{.}
\end{equation}
The first (second) summation on the right hand side adds up transition amplitudes of flip-flop processes that transform population to (from) state $\left| j \right\rangle$. From the solution of the above systems of linear equations one can obtain $da_{0l} $ and $da_{il} $.

 It is important to notice that, $da_{0l}$ and $da_{il}$ are not always uniquely defined. Altogether ${d_0 \left( d_0 - 1 \right)}/{2}$ number of transition amplitudes can be nonzero simultaneously in a given cluster system $c_i$. On the other hand, maximally $d_0 - 1$ independent linear equations can be defined from the diagonal elements of $\text{Tr}_i d\varrho_{c_i}^{h\mathcal{E}}$ in the cluster system. Therefore, $da_{il}$ are unambiguously defined only for $d_0 = 2$. We note, however, that the number of required $C_l$ operators may be reduced by invoking system dependent physical considerations. 
It is often the case that only $\Delta q = \pm 1$ spin flip-flop processes are possible in a basis defined by $q$ quantum number. When the required number of $C_l$ operators is either equal to or less than $ 2\left( d_0-1 \right)$ all $da_{il}$ amplitudes can be uniquely defined. 
Finally, in Appendix~\ref{app:ail_det} we discuss how to determine $da_{il}$ in cases when the number of possible non-zero $da_{il}$ amplitudes is larger than $d_0-1$.

Having all $da_{0l}$ and $da_{il}$ transition amplitudes defined, in step (vii) of the propagation cycle we compute 
\begin{equation}
 db_{0l}  = \dot{b}_{0l}dt = \sum_{i=1}^{n} \Delta d{a}_{il} 
\end{equation}
and 
\begin{equation}
d{b}_{il} = \dot{b}_{il}dt  = \sum_{j=1, j \neq i}^{n} \Delta d{a}_{jl}   \text{,}
\end{equation}
where
\begin{equation}
\Delta da_{il} = da_{il} - da_{0l} \text{.}
\end{equation}

In step (viii), we determine the variation of the density matrices due to the extended Lindbladian defined in  Eq.~(\ref{eq:ext_linbd_0}) and Eq.~(\ref{eq:ext_linbd_i}) as
\begin{equation}
d\varrho_{c_0}^{\mathcal{L}} \!  \left( t \right) = \mathcal{L}_{c_0} \! \left( \left\lbrace db_{0l} \right\rbrace , \left\lbrace C_{0l} \right\rbrace ; \varrho_{c_0} \! \!  \left( t \right)  \right) 
\end{equation}
and 
\begin{equation}
d\varrho_{c_i}^{\mathcal{L}} \!  \left( t \right) =  \mathcal{L}_{c_i} \! \left( \left\lbrace db_{il} \right\rbrace , \left\lbrace C_{il} \right\rbrace ; \varrho_{c_i} \! \!  \left( t \right) \right)  \text{.}
\end{equation}

In step (ix) of the propagation cycle, the cluster density matrices at $t+dt$ are determined as
\begin{equation}
\varrho_{c_0} \! \!  \left( t + dt \right) =  \varrho_{c_0} \! \!  \left( t \right) + d\varrho_{c_0}^{h\mathcal{E}} \!  \left( t \right) + d\varrho_{c_0}^{\mathcal{L}} \!  \left( t \right)
\end{equation}
and
\begin{equation}
\varrho_{c_i} \! \!  \left( t + dt \right) =  \varrho_{c_i} \! \!  \left( t \right) + d\varrho_{c_i}^{h\mathcal{E}} \!  \left( t \right) + d\varrho_{c_i}^{\mathcal{L}} \!  \left( t \right) \text{.}
\end{equation}

Finally, repetition of this procedure by substituting $\varrho_{c_0} \!   \left( t  \right)$ and  $\varrho_{c_i} \!   \left( t  \right)$ by $\varrho_{c_0} \!   \left( t + dt \right)$ and  $\varrho_{c_i} \!  \left( t + dt \right)$ allows one to  simulate the dynamics of the many spin system $\mathcal{S}$.

 \section{Spin dynamics in cluster approximation}
 \label{sec:dyn}

\begin{figure}[h!]
	\includegraphics[width=0.450\columnwidth]{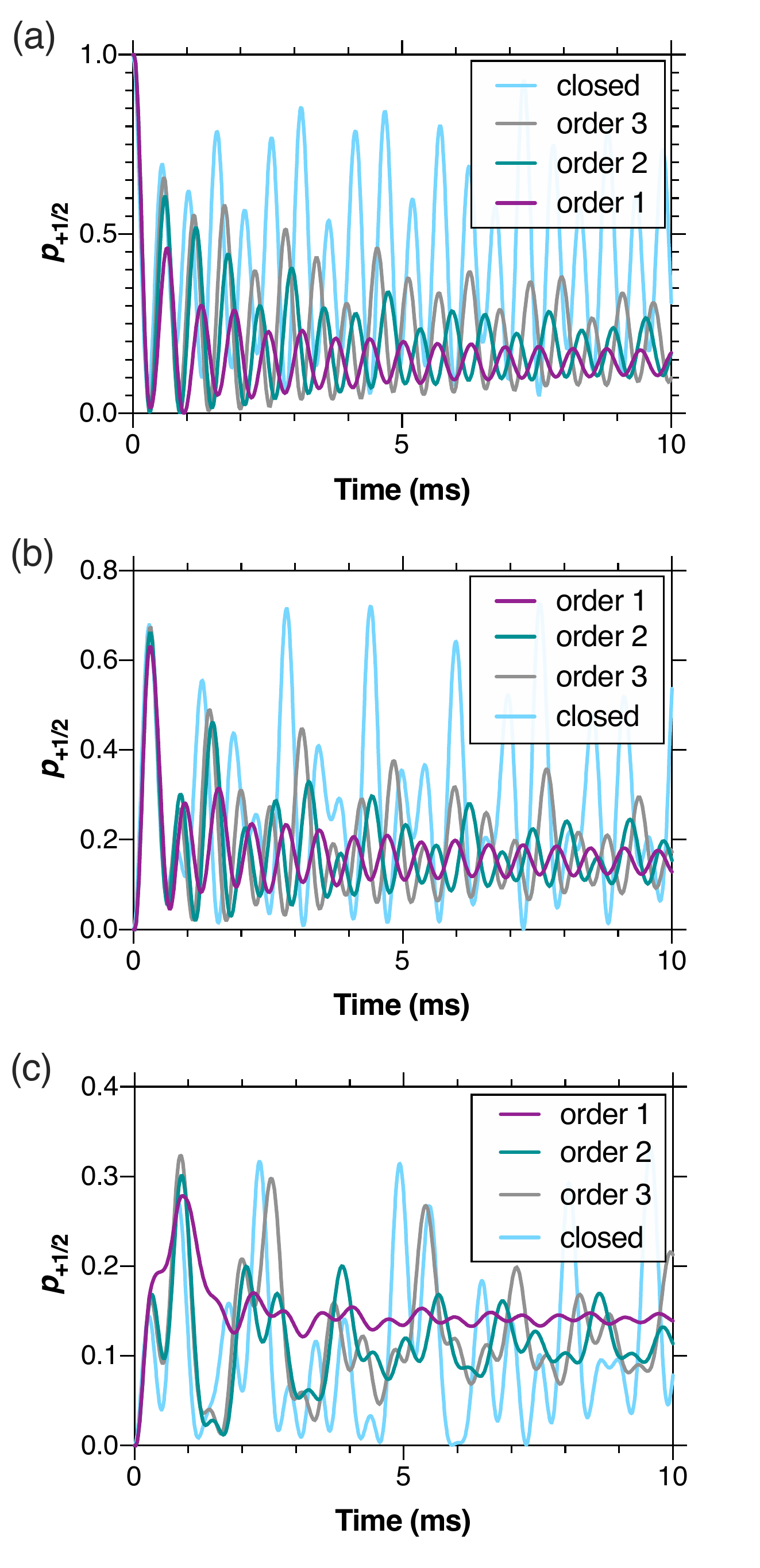}
	\caption{ Comparison of exact and approximate time evolution. (a), (b), and (c) show the time evolution of the  projection $\left\langle \left. +1/2 \right| m_S \right\rangle$, of the central spin, the strongest coupled environmental spin, and the second strongest coupled environmental spin in a central spin arrangement of seven (1+6) spin-1/2, respectively. Light blue and gray, teal, and plum curves depict the exact time evolution of the closed system and the time evolution obtained in order 3, 2, and 1 cluster approximations.  }
	\label{fig:eva-noT2}  
\end{figure}

In order to elaborate on the properties of the proposed method, first we study the time evolution of an exemplary spin system obtained from different level of approximation and exact propagation. The considered system consists of seven spin-1/2 spins in a central spin arrangement. We write the Hamiltonian of the system as
\begin{equation}
H_0 = B S_z + \sum_{i=1}^{6} A_i \mathbf{S} \mathbf{I}_i  \text{,}
\end{equation}
where $\mathbf{S}$ and $\mathbf{I}_i$ are  spin operator vectors of the central and  environmental spins, $S_z$ is the spin $z$ operator of the central spin, and $A_i = 1/i$~MHz are the coupling constants for $i$ goes from 1 to 6. $B$ is set either to zero or to 100~MHz that represent either strong or week coupling limit, respectively. At $t = 0$ the central spin is polarized in the $\left| +1/2 \right\rangle$ state, while the environmental spins are in the $\left| -1/2 \right\rangle$ state. 

Time evolution of selected spins, such as the central spin and the two strongest coupled environmental spins, of the strongly coupled central spin model is depicted in Fig.~\ref{fig:eva-noT2}. Exact propagation of the closed system shows coherent oscillations. We also see coherent oscillations in all approximate solutions, however, the amplitude of these oscillations decays. This is due to the neglect of the intra-spin bath entanglement and the Lindbladian drive of the different cluster systems. Timescale of the approximation caused decoherence extends, however, with increasing cluster approximation orders. In addition, fine structure of the coherent beatings is also improved in higher order approximations, see for example Fig.~\ref{fig:eva-noT2}(c). 

\begin{figure}[h!]
	\includegraphics[width=1.0\columnwidth]{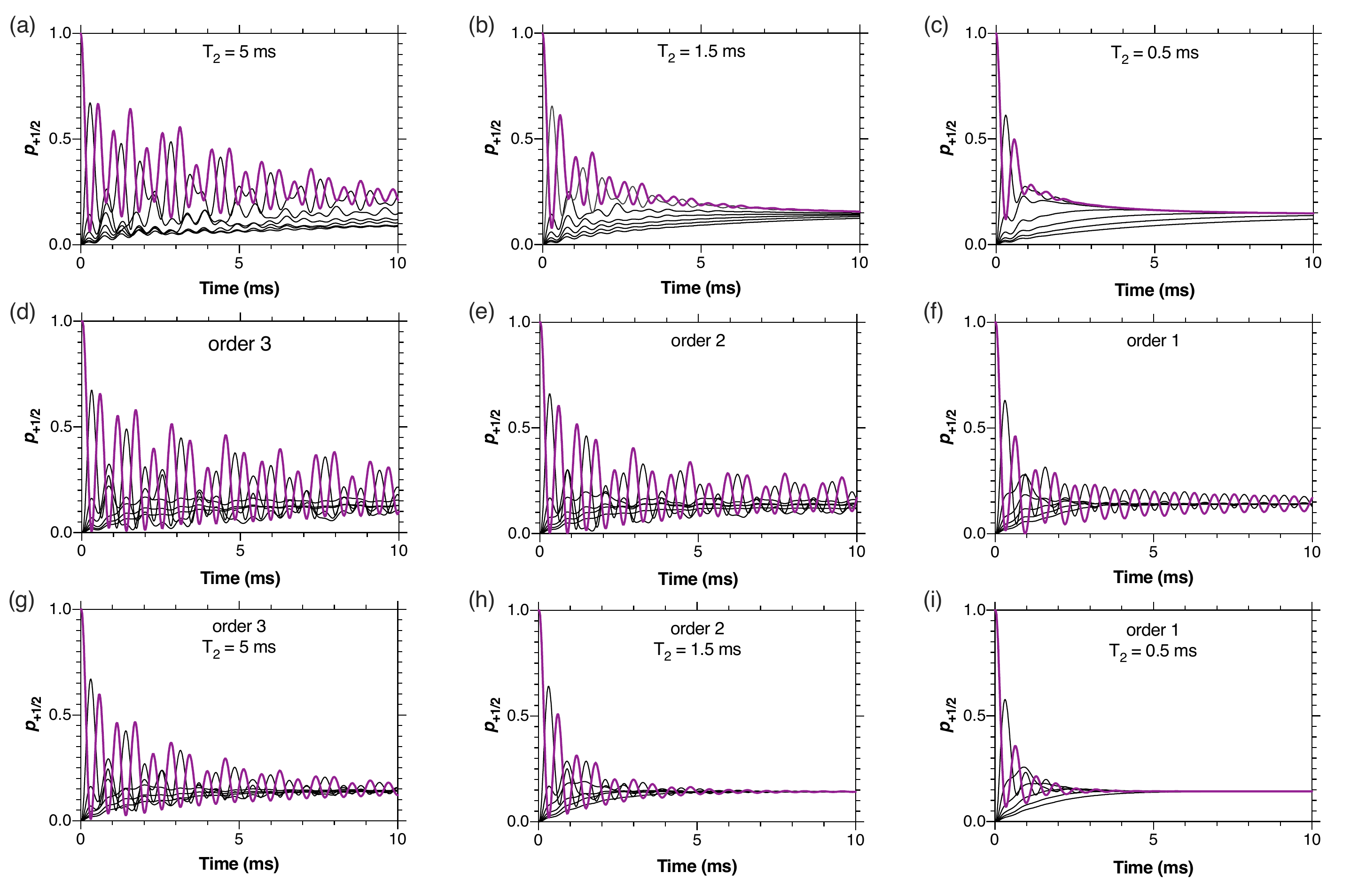}
	\caption{ Time evolution and dephasing. (a), (b), and (c) show Lindbladian time evolution of the 7-spin central spin system assuming dephasing of the environmental spins on time scale T$_{2}$. (d), (e), and (f) depict the time evolution obtained from order 3, 2, and 1 approximation, respectively, with no additional Markovian dephasing, while (g), (h), and (i) depict the time evolution of the various cluster approximations including Markovian dephasing of time scale T$_{2}$.  In all cases plum and thin solid lines represent the time evolution of the central spin and the six environmental spins, respectively}
	\label{fig:eva-T2}  
\end{figure}

To further investigate the nature of the spurious decoherence, we compare the time evolution of the model system subject to Markovian dephasing of the environmental spins, Fig.~\ref{fig:eva-T2}(a)-(c), with the cluster approximation method either excluding, Fig.~\ref{fig:eva-T2}(d)-(f), or including, Fig.~\ref{fig:eva-T2}(g)-(i), additional Markovian dephasing of the environmental spins. As can be seen in Fig.~\ref{fig:eva-T2}(a)-(c) dephasing of the environment does give rise to decay of the coherent oscillations of the central and the environmental spins, similarly to the approximate solution seen in Fig.~\ref{fig:eva-T2}(d)-(f) for different orders. There are two differences between the characteristics of the decaying curves. 1) Coherent oscillations decay exponentially due to Markovian decoherence, while the envelop of the decaying curves in the cluster approximation follows more like a stretched exponential $\exp \left( - t^{\alpha} \right)$, where $\alpha < 1$. It is worth mentioning that after combining Markovian decoherence with different order of cluster approximation the decaying curves  resemble exponential decaying coherent oscillations. 2) Polarization of weakly coupled environmental spins increases faster than in the Markovian case. All of the curves shown in Fig.~\ref{fig:eva-T2} preserve the net spin quantum number of the model system, therefore decaying curves of all the spins approach the value of $1/7$ that corresponds to equal polarization of the spins.

\begin{figure}[h!]
	\includegraphics[width=0.60\columnwidth]{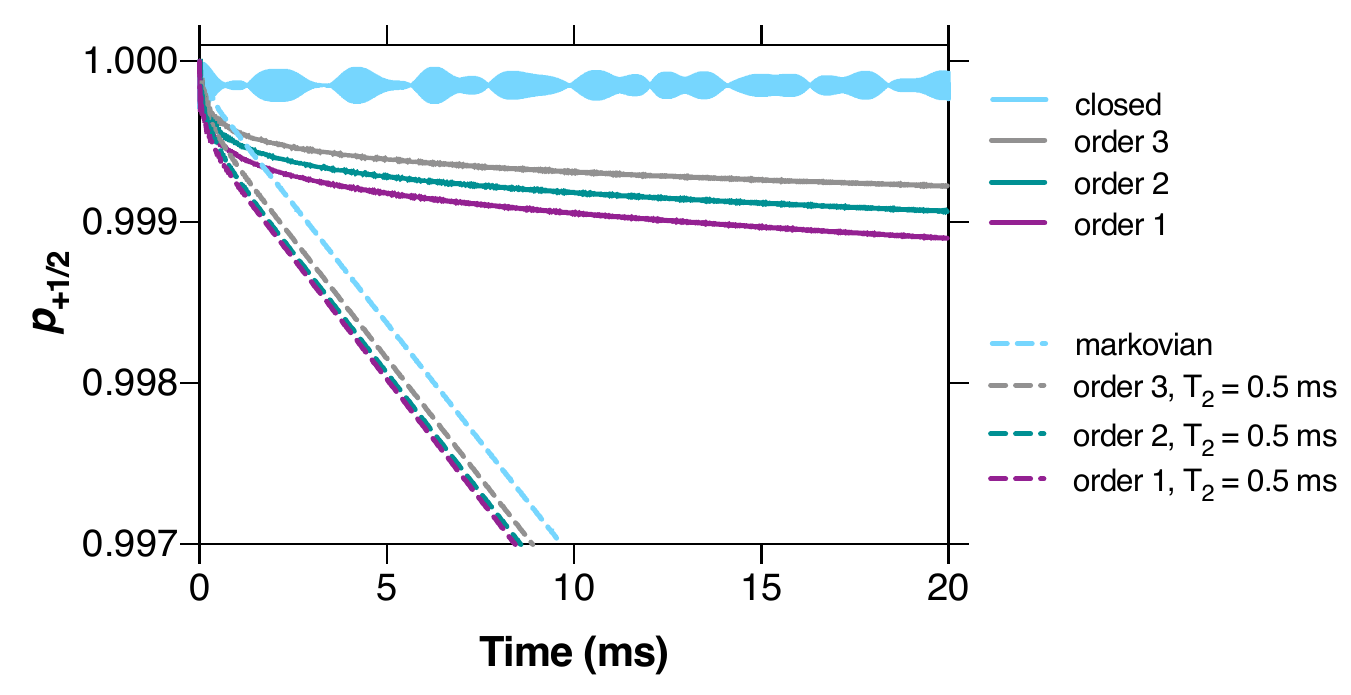}
	\caption{ Time evolution of a decoupled spin system. Solid light blue curve shows time evolution of a closed central spin system, where the splitting of the central spin's energy levels is two orders of magnitude larger than the strongest environmental coupling. Solid gray, teal, and plum curves show time evolution the same system in order 3, 2, and 1 cluster approximations of the central spin. Dashed curves depict time evolution of open systems where Markovian dephasing of the environmental spins are assumed. }
	\label{fig:eva-100MHz}  
\end{figure}

Finally, we investigate weakly coupled cases when the couplings to the environmental spins are largely suppressed by a $B = 100$~MHz splitting introduced between the energy levels of the central spin. Fig.~\ref{fig:eva-100MHz} summarizes our findings. Exact time evolution of the closed system show very fast oscillations modulated by lower frequency oscillations. It is important to notice that the curve does not decay. In cluster approximations, the curves miss the fast coherent oscillations, except the very beginning, and decay stretched exponentially. In both cases exponential decay of the initial high polarization can be induced by additional Markovian dephasing of the spin bath.

 \section{Case study: T$_1$ of NV center in diamond}
 \label{sec:sum_res}

In this section, we computationally demonstrate through the example of NV center in diamond that the above described method can account for spin relaxation processes in different spin environments at various external fields. First, we provide spin Hamiltonians for the considered systems, then we describe the details of our \emph{ab initio} density funtional theory calculations used to parameterize NV center-nuclear spin bath interactions. In the subsequent sections we study different spin bath induced relaxation processes of an NV center's spin polarization.

\subsection{ Background and methodology}

\subsubsection{Spin Hamiltonian}

We study NV center-spin bath coupled systems. In particular, we consider P1 center (neutral substitutional nitrogen atom with spin-1/2 ground state), NV center, and $^{13}$C nuclear spin reservoirs interacting with the central NV center ($s_0$). For simplicity, we ignore NV centers' nitrogen nuclear spin that  gives rise only to a fine structure at the ground state level anti crossing (GSLAC)\citep{IvadyDNP2015}. The nitrogen nuclear spin of the P1 center is, however, taken into consideration due to its strong, $\mathcal{O} \! \left( 100 \text{ MHz} \right)$ hyperfine coupling. The spin Hamiltonian $h^{\text{cNV}}$ of the central spin, $h^{\text{P1}}_i $ of P1 centers, $h^{\text{eNV}}_i$ of environmental NV centers, and $h^{\text{13C}}_i $ of $^{13}$C nuclear spins can be written as,
\begin{equation}\label{eq:h0NV}
h^{\text{cNV}} = D \left( S_z^2 - \frac{2}{3} \right ) + g_e \mu_B B_z S_z + d_{\parallel } S_z^2 + d_{\perp} \left( \left\lbrace S_x, S_z \right\rbrace + \left\lbrace S_y, S_z \right\rbrace + \left\lbrace S_x, S_y \right\rbrace + S_y^2 - S_x^2 \right) \text{,}
\end{equation}
where $S_z$ is the spin $z$ operator defined in a coordinate system with $z$-axis parallel to the NV axis, $D = 2.870$~GHz \citep{Gruber:Science1997} is the zero field splitting, $g_e$ is the electron $g$-factor, $\mu_{B}$ is the Bohr magneton, and $d_{\parallel}$ and  $d_{\perp}$ account for parallel and perpendicular strain coupling\citep{UdvarhelyiStrain2018}, respectively,
\begin{equation}\label{eq:hiP1}
h^{\text{P1}}_i = g_e \mu_B B_z S_{i,z}  + \widetilde{S}_i \widetilde{A} \widetilde{J}_{i} + P \left( \widetilde{J}_{i,z}^2 - \frac{2}{3} \right ) + g_{14N} \mu_N B_z J_{i, z}  \text{,}
\end{equation}
where variables with tilde symbol are defined in a coordinate system with $z$-axis parallel to the C$_{3v}$ axis of the Jahn-Teller distorted configuration of P1 center,  $J_{i}$ is the $^{14}$N nuclear spin operator, $P = 5.01$ is the quadrupole splitting for which we use the value of the NV center \citep{Felton2008}, $A$ is the hyperfine tensor whose diagonal elements, $A_{zz} = 114$~MHz and $A_{xx} = A_{yy} = 81$~MHz, are determined by our first principles electronic structure calculations, see below, $g_{14N} = 0.4038$ is the nuclear g-factor of the spin-1 $^{14}$N nucleus, $\mu_{N}$ is the nuclear magneton,
\begin{equation}\label{eq:hiNV}
h^{\text{eNV}}_i = D \left( \widetilde{S}_{i,z}^2 - \frac{2}{3} \right ) + g_e \mu_B B_z S_{i,z} + d_{\parallel }  \widetilde{S}_z^2 + d_{\perp} \left( \left\lbrace  \widetilde{S}_x,  \widetilde{S}_z \right\rbrace + \left\lbrace  \widetilde{S}_y,  \widetilde{S}_z \right\rbrace + \left\lbrace  \widetilde{S}_x,  \widetilde{S}_y \right\rbrace +  \widetilde{S}_y^2 -  \widetilde{S}_x^2 \right) \text{,}
\end{equation}
 where  $\widetilde{S}_i$ is the spin operator defined in a coordinate system with $z$ axis parallel to the symmetry axis of the environmental NV center, and
\begin{equation}\label{eq:hiC13}
h^{\text{13C}}_i = g_{13C} \mu_N B_z I_{i, z}  \text{,}
\end{equation}
where $I_{i }$ and $g_{13C} = 1.4048$ are the nuclear spin operator and the nuclear g-factor of the spin-1/2 $^{13}$C nucleus, respectively.

Coupling tensors between the central NV center's spin and electron spin bath specimens, such as P1 centers and environmental NV centers, are obtained by neglecting spatial distribution of the spin densities through the dipole-dipole interaction Hamiltonian,
\begin{equation}
h^{SS}_{0i} = -\frac{\mu_0}{4 \pi} \frac{g^2 \mu_{B}^{2}}{ \left| r_{0i} \right|^3} \left( 3 \left(S_0 r_{0i} \right) \left( S_i r_{0i} \right) - S_0 S_i \right) \text{,}
\end{equation}
where $S_0$ and $S_i$ are spin operators of the central spin and the coupled spin, respectively, $\mu_0$ is the vacuum permeability, and $r_{0i}$ is a vector pointing from the central spin to the coupled spin. When a nuclear spin bath is considered, NV center-nuclear spin couplings are described by the hyperfine interaction Hamiltonian,
\begin{equation}
h^{SI}_{0i} = S_0 A_i I_i\text{,}
\end{equation}
where $I_i$ and $A_i$ are the nuclear spin operator and the hyperfine coupling tensor in cluster system $j$, respectively. 

The following cluster Hamiltonians are used to model different spin environments,
\begin{eqnarray}
h_{c0}^{\text{P1}} =  h_{c0}^{\text{NV}} = h_{c0}^{\text{13C}} = h^{\text{cNV}} \text{,} \\
h_{ci}^{\text{P1}} = h^{\text{cNV}} + h^{\text{P1}}_i + h^{SS}_{0i} \text{,} \\
h_{ci}^{\text{NV}} = h^{\text{cNV}} + h^{\text{eNV}}_i + h^{SS}_{0i} \text{,} \\
h_{ci}^{\text{13C}} = h^{\text{cNV}} +  h^{\text{13C}}_i +  h^{SI}_{0i} \text{.} 
\end{eqnarray}

\subsubsection{First principles electronic structure calculations}

Hyperfine coupling tensors are key quantities when a $^{13}$C nuclear spin bath is considered. We use first principles Density Functional Theory (DFT) electronic structure and subsequent hyperfine tensor calculations to obtain relevant coupling tensors of  NV center and P1 center spin systems in diamond. In our DFT calculations we use a 1728 atom supercell, HSE06 hybrid functional\citep{HSE03}, PAW core potentials\citep{PAW}, and plane wave basis set of 420~eV as implemented in VASP\citep{VASP2}.

It is possible to calculate hyperfine interaction with high accuracy\citep{Szasz13} for atomic sites in close vicinity of the NV center, however, for farther sites the hyperfine interaction suffers from considerable finite size effects in supercell methods \citep{Davidsson2018,IvadyNPJ2018}. To overcome this issue we utilize a real space grid combined with the PAW method to calculate  hyperfine tensors. The Fermi contact term, dipole-dipole interaction within the PAW sphere, and core polarization corrections are calculated within the PAW formalism\citep{Szasz13} from the convergent spin density. The dipolar hyperfine contribution from spin density localized outside the PAW sphere is calculated by using a uniform real space grid. This procedure allows us to obtain hyperfine coupling tensors excluding effects from periodic replicas of the spin density due to the periodic boundary condition. Additionally, we can calculate hyperfine tensors for atomic sites outside the boundaries of the supercell by neglecting Fermi contact interactions in that region.

 \subsection{Results}

In the following computational example, we study the NV center's longitudinal spin relaxation in different spin environments over a wide range of external magnetic fields and strain. In the simulations, we neglect spin-orbit and phonon assisted decay processes. The former effect is negligible in the ground state of the NV center, while the latter approximation is valid at low temperatures (below $\approx$50~K). 

First, we investigate spin relaxation due to P1 center spin environment. In the simulations, we considered an ensemble of 50 randomly generated configurations of 31 P1 centers. The concentration of the P1 center spin bath is set to 50 ppm, which corresponds sample S2 in 
Ref.~[\onlinecite{Jarmola2012}].
Except for $c_0$, each $c_i$ cluster system include the central NV center ($s_0$) and one P1 center ($s_i$) from the environment. The interaction between the environmental spins is neglected. We note that P1 centers in diamond can have different orientations depending on whether their symmetry axis is parallel or 109$^{\circ}$ aligned to the axis of the external magnetic field and the central NV center. Relaxation effects due to differently oriented electron spin defects are studied separately. The density matrices of the cluster systems at $t = 0 $ describe the central NV center polarized in $m_{\text{S}} = 0$ and a non-polarized P1 center. Four $C_l$ Lindblad operators are defined to account for $\left| 0 \right\rangle \leftrightarrow  \left| \pm 1\right\rangle$ transitions of the central spin. Spin dynamics simulations model the time evolution of the coupled system over 1~ms time period, during which the central spin slowly looses its polarization. The decay time $T_1$ is obtained by fitting an experiential function to the resultant polarization curve of the central NV center.

\begin{figure}[h!]
	\includegraphics[width=0.95\columnwidth]{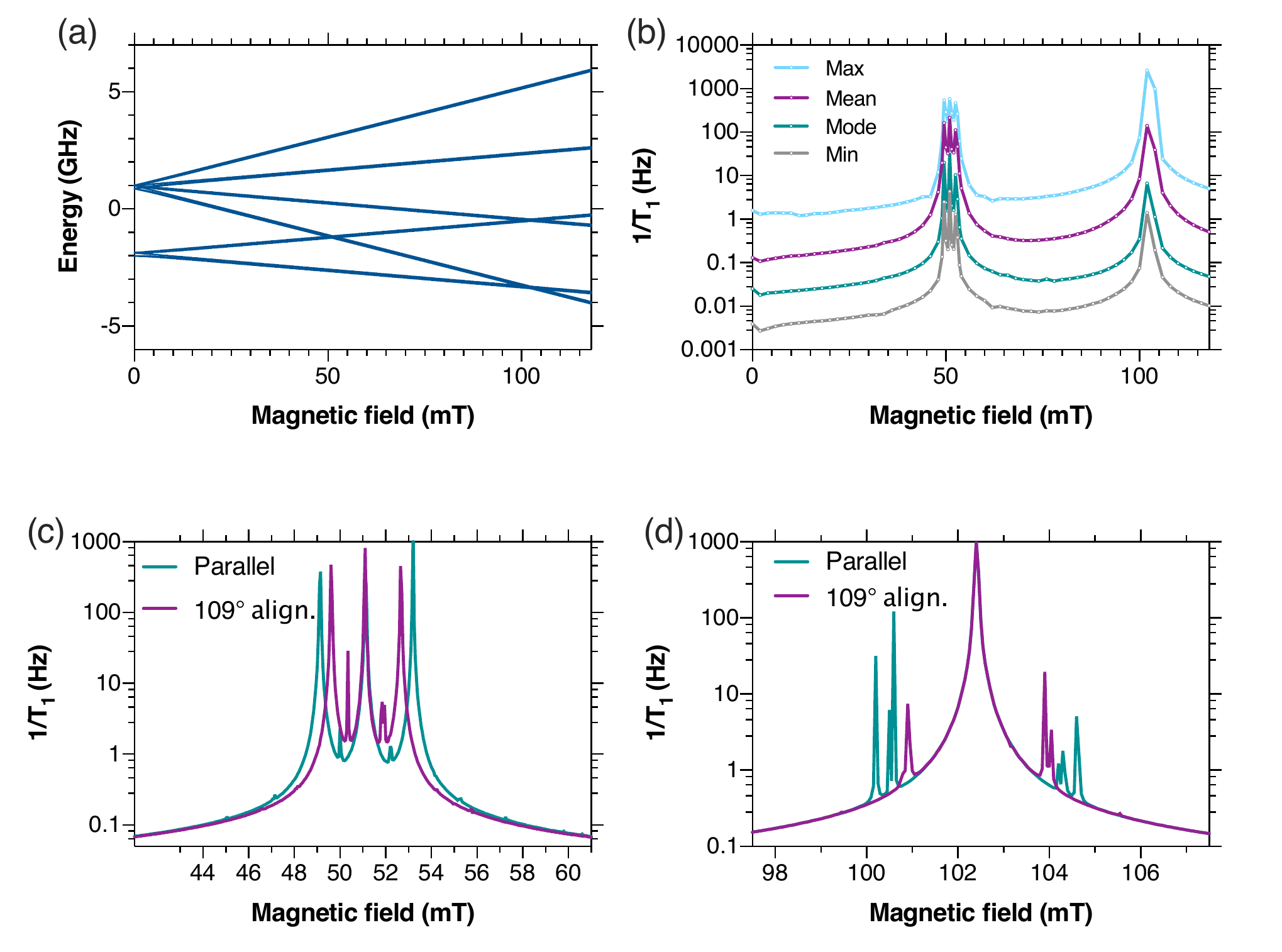}
	\caption{ Spin relaxation in P1 center environment. (a) Energy level structure of NV-P1 two electron spin system as a function of external magnetic field.  (b) Magnetic field dependence of the spin relaxation rate ($1/T_1$). Light blue, plum, teal, and gray curves show the largest (Max), average (Mean), most probable (Mode), and lowest (Min) relaxation rates obtained in an ensemble of 50 randomly generated arrangements of 31 P1 centers that corresponds to 50~ppm defect concentration on average.  (c)  and  (d) depict the fine structure of spin relaxation rate at 51~mT and 102~mT, respectively, for a representative P1 center arrangement. Teal and plum curves show the contributions of parallel and 109$^{\circ}$ aligned P1 centers of equal concentration. }
	\label{fig:P1}  
\end{figure}

Spin relaxation rate ($1/T_1$) as a function of the external magnetic field is depicted in Fig.~\ref{fig:P1}(b). Note that the distribution of the relaxation rates over the considered ensemble is highly asymmetric, meaning that there is a low but non-zero probability of finding centers with very large relaxation rates. Such a distribution cannot be faithfully  characterized by the usual statistical quantities, such as mean and standard deviation, therefore in Fig.~\ref{fig:P1}(b), we provide additional quantities to properly describe the relaxation rate distribution. We depict the relaxation rate of configurations with the lowest (Min) and the highest (Max) relaxation rates, as well as, the mean (Mean) and the mode (Mode) of the distribution. Note that the mean and mode relaxation rates are different due to the asymmetric distribution, implying that the average $T_1$ time can, in general, be shorter than the most probable $T_1$ time of individual centers in an ensemble.

The simulations reveal two magnetic field values, 51~mT and 102~mT, where enhanced spin relaxation takes place. By looking at the energy level structure of NV-P1 coupled system depicted in Fig.~\ref{fig:P1}(a), the relaxation peaks can be assigned to level crossings. Reduction of energy gaps enhances spin flip-flop rates, which is captured by the simulations. We note that at 0~mT, no enhanced relaxation can be observed despite the crossing of the levels at this field. This is due to the fact that the spin-1 NV center exhibits a large zero-field splitting, while electron spin sublevels of the P1 center are degenerate at zero magnetic field. Therefore, couplings are efficiently suppressed. 

The relaxation peak at 51~mT exhibits a fine structure not fully resolvable in Fig.~\ref{fig:P1}(b).  In Fig.~\ref{fig:P1}(c), we depict the relaxation rate of a representative spin bath configuration, including either parallel or 109$^{\circ}$ aligned P1 centers. In both cases a five-peak fine structure can be seen with different spacings due to the different orientation of the hyperfine principal axis in the two cases. Related structures were recently observed in electron paramagnetic resonance (EPR)\citep{Wood2016}, photo luminescence (PL)\citep{Armstrong2010,Hall2016,Wickenbrock2016}, and NMR measurements\citep{FischerPRL2013,Wunderlich2017}. 

A different fine structure is obtained at 102~mT, see Fig.~\ref{fig:P1}(d). The peaks at 51~mT are due to spin flip-flop interactions between the NV center and P1 centers, while the central peak at 102~mT is due to the precession of the NV spin in the transverse magnetic field of the P1 centers, and the side peaks near 102~mT are due to three spin processes assisted by the $^{14}$N nuclear spin of the P1 center. Related PL signatures were recently reported in 
Ref.~[\onlinecite{Wickenbrock2016}].

The main approximation of the methodology proposed in this article is the neglect of entanglement between environmental spins. For P1 center spin bath we obtained $T_1 > 1$~ms for most of the magnetic field values considered in the simulations. As the $T_2^{\text{P1}}$ time of the P1 centers at 50~ppm is expectedly much shorter than 1~ms, the relation  $T_2^{\text{P1}} << T_1^{\text{NV}}$ is satisfied. This validates the approximation of non-entangled spin bath.

\begin{figure}[h!]
	\includegraphics[width=0.85\columnwidth]{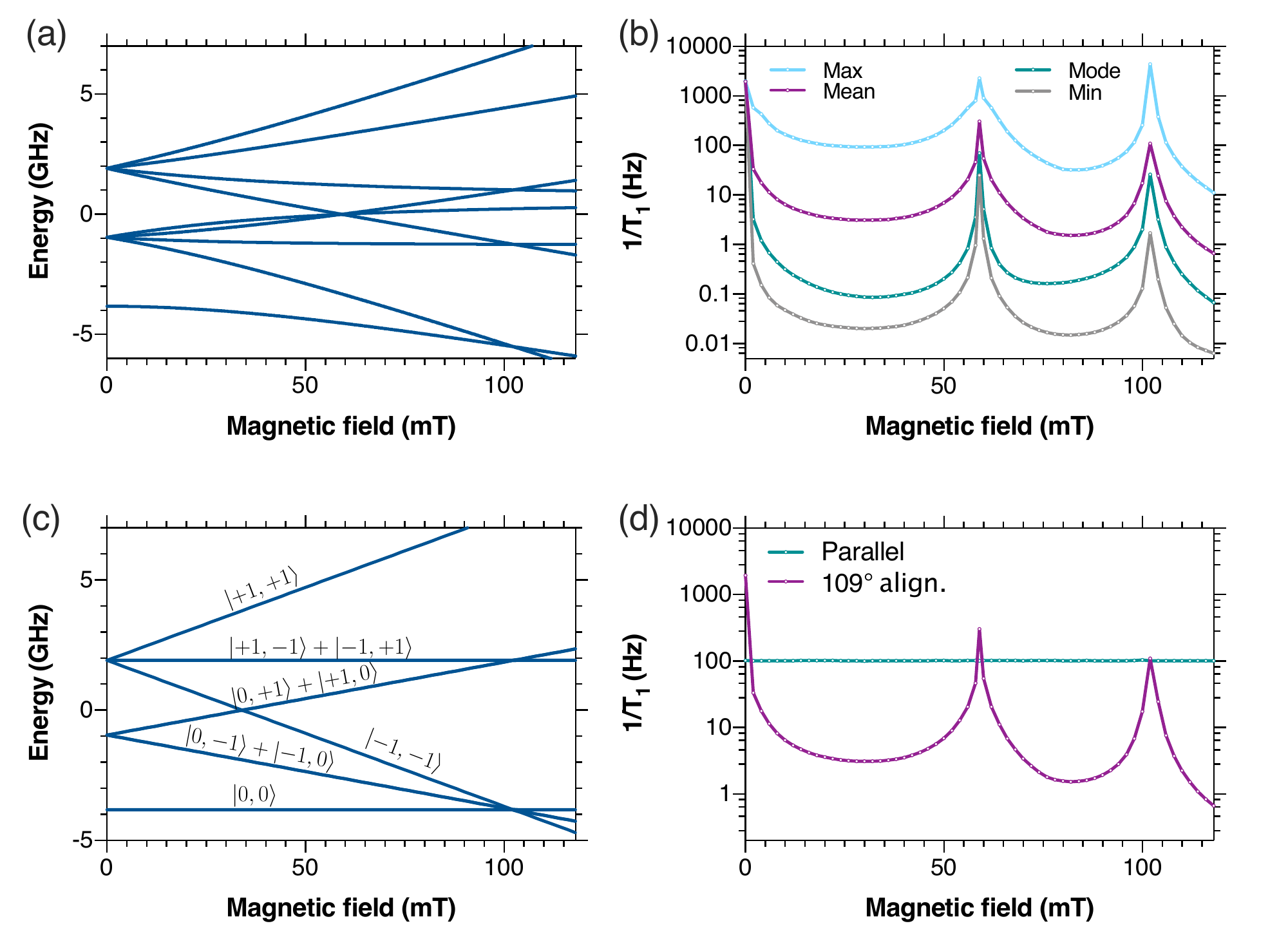}
	\caption{Spin relaxation  in NV center environment. (a) Energy level structure of central NV-109$^{\circ}$ aligned NV two spin system as a function of external magnetic field.  (b) Magnetic field dependence of the corresponding spin relaxation rate ($1/T_1$). Light blue, plum, teal, and gray curves show the largest (Max), average (Mean), most probable (Mode), and lowest (Min) relaxation rates obtained in an ensemble of 50 randomly generated arrangements of 31, 109$^{\circ}$ aligned NV centers that corresponds to 8~ppm defect concentration on average.  (c) Energy level structure of central NV center-parallel NV center system as a function of external magnetic field. (d) Magnetic field dependence of ensemble averaged spin relaxation rate due to parallel (teal) and 109$^{\circ}$ aligned (plum) NV centers. 
	}
	\label{fig:NV}  
\end{figure}

Next, we investigate the magnetic field and strain dependence of the spin relaxation rate of a central NV center interacting with a number of environmental NV centers. Settings for the simulations are the same as for P1 center environment, except the defect concentration, which is set to 12~ppm,
in accordance with sample S2 in 
Ref.~[\onlinecite{Jarmola2012}],
 and the initialization of the environmental NV centers, where we set 90\%  polarization in the $m_{\text{S}} = 0$ state.

The energy level structure and the corresponding theoretical spin relaxation rate of a central NV center in 109$^{\circ}$ aligned NV center environment are depicted in Fig.~\ref{fig:NV}(a) and (b), respectively. We obtain highly anisotropic distributions for the relaxation rates characterized by the minimal, maximal, mean, and mode values in Fig.~\ref{fig:NV}(b). Three relaxation peaks can be found in the investigated magnetic field interval at 0~mT, 59~mT, and 102~mT.  Related PL features at 59~mT were reported in experiment.\citep{Armstrong2010,Wickenbrock2016} The peaks correspond to crossings between the energy levels of the coupled two NV center systems depicted in Fig.~\ref{fig:NV}(a). Since the central NV center and the environmental NV centers exhibit the same zero-field splitting, spin states can be mixed at zero magnetic field that gives rise to a relaxation peak, in contrast to the P1 center environment.

A magnetic field oriented NV center environment gives rise to a distinct relaxation pattern, see Fig.~\ref{fig:NV}(d). The obtained high and constant relaxation rate can be explained by looking at the energy level structure of mutually aligned NV center pair system in Fig.~\ref{fig:NV}(c). One can see that two pairs of energy levels, correspond to $\left| 0, -1\right\rangle$ and $\left| -1, 0 \right\rangle$ states and  $\left| 0, +1\right\rangle$ and  $\left| +1, 0 \right\rangle$ states, are degenerate irrespective of the magnetic field. This is due to the identical Hamiltonian of the two centers. The degenerate states can be mixed by the dipole-dipole coupling which gives rise to a constant very high relaxation rate. 
We note that this high relaxation rate can be substantially reduced in experiment due to two effects. (i) The relaxation rate is depends linearly on the polarization difference between the central NV center and the environmental NV centers. In our simulations we set a 10\% difference, which may be higher than in sample upon measurement. (ii) In the simulation the states are degenerate due to the identical level structures of the two centers, however, magnetic field and strain inhomogeneities may make the centers distinguishable. 

\begin{figure}[h!]
	\includegraphics[width=0.85\columnwidth]{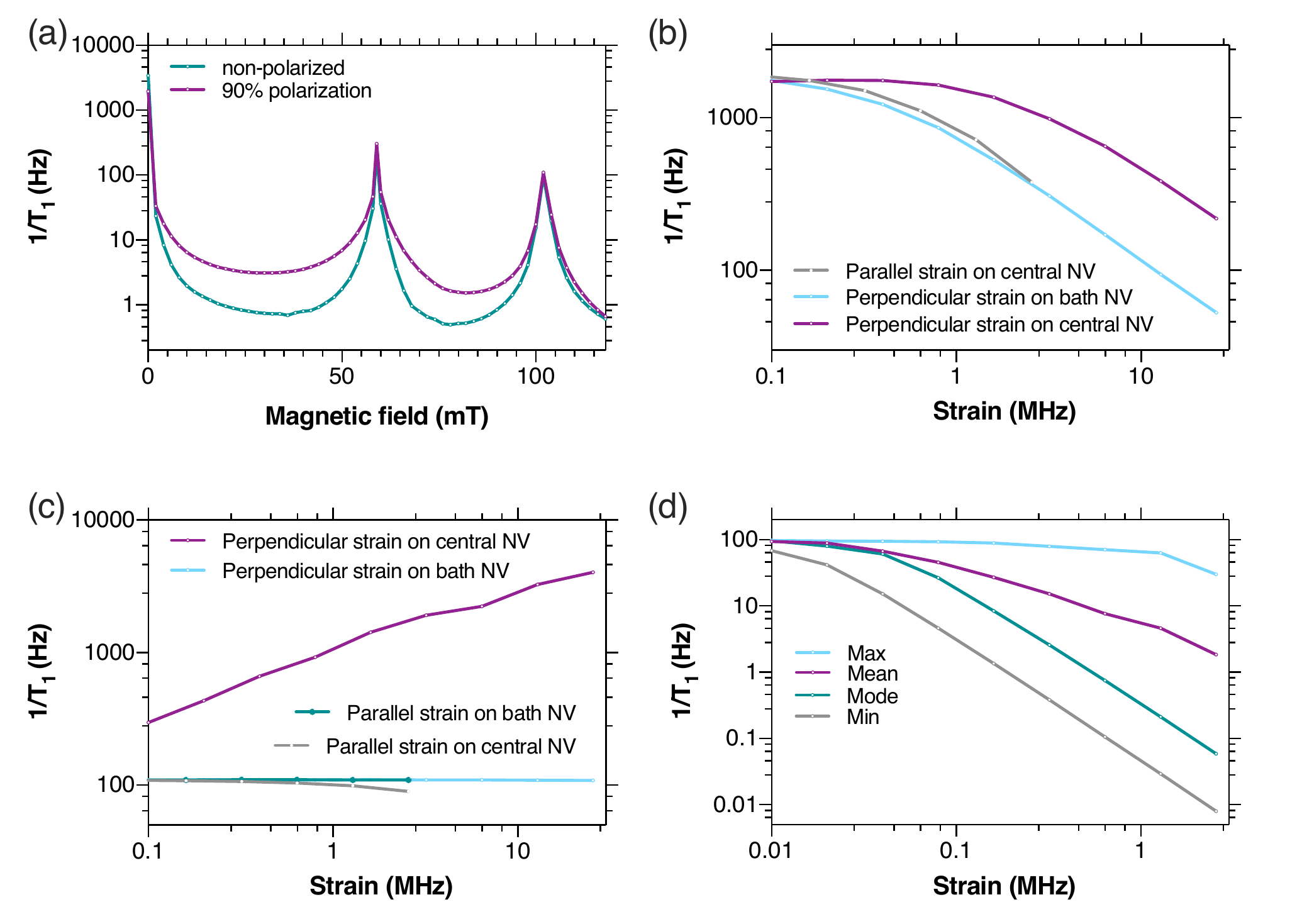}
	\caption{Bath polarization and strain dependence. (a) Magnetic field dependent relaxation rate of a central NV center interacting with non-polarized (teal) and 90\% polarized (plum) 109$^\circ$ aligned NV center bath of 8 ppm concentration.  (b) and (c) depict strain dependence of the relaxation rate of NV center in 109$^\circ$ aligned NV center environment at $B = 0$ and $B = 102$~mT, respectively. Effects of different strain components are considered separately. (d) Strain dependence of the spin relaxation rate due to parallel NV environment of 4~ppm concentration.
	}
	\label{fig:NVstrain}  
\end{figure}

In order to investigate these effects in NV center-NV bath systems, we actuate parallel and perpendicular strain on the central spin and environmental NV centers of parallel and 109$^{\circ}$ alignment. In Fig.~\ref{fig:NVstrain}(a), polarization dependence of the 109$^{\circ}$ aligned NV center bath induced spin relaxation rate is depicted. We find that the relaxation rate is approximately a factor of three larger in the case of the polarized, 90\%  in $m_{S} = 0$ state with 109$^{\circ}$ aligned quantization axis, NV center bath. Due to the optical pumping, polarization of the NV bath is expected, however,  at magnetic field strengths with enhanced coupling to other spin species, e.g.\  at $B = 59$~mT, low polarization is more probable.  

Strain dependence of the relaxation rate at specific magnetic fields is depicted in Fig.~\ref{fig:NVstrain}(b)-(c) and Fig.~\ref{fig:NVstrain}(d) for 109$^{\circ}$ aligned and parallel NV center environments, respectively. At $B = 0$, both parallel and perpendicular strain applied on central and environmental NV centers effectively lower the relaxation rate due to the opening of small gaps between the degenerate states. Note that the coupling of the NV center to perpendicular strain is an order of magnitude larger than the coupling to parallel strain, thus the range of considered strain field is larger in the former case. Note furthermore that, similar but reduced effects can be found at $B=59$~mT, not shown. At $B=102$~mT we see, however, distinct behavior. Relaxation rate appears insensitive to parallel strain to a large extent, when applied on the central NV center and to both parallel and perpendicular strain applied on environmental NV center. On the other hand perpendicular strain applied on the central NV center mixes the spin states efficiently\citep{UdvarhelyiStrain2018} which gives rise to a prominent increase of the relaxation rate. Relaxation rate distribution of NV centers in parallel aligned NV center environment is characterized in Fig.~\ref{fig:NVstrain}(d). It is apparent from the figure that the strain shift reduces the relaxation rate substantially. This effect, however, vary considerably with the spin bath configurations. When the central spin-environment couplings are weak, even a small strain shifts can induce large reductions in the rates.

Similar to the P1 center environment, the condition $T_2^{\text{NV}} << T_1^{\text{NV}} $ is expectedly satisfied in the modeled sample. Therefore, the approximations of the applied method hold. 

\begin{figure}[h!]
	\includegraphics[width=0.95\columnwidth]{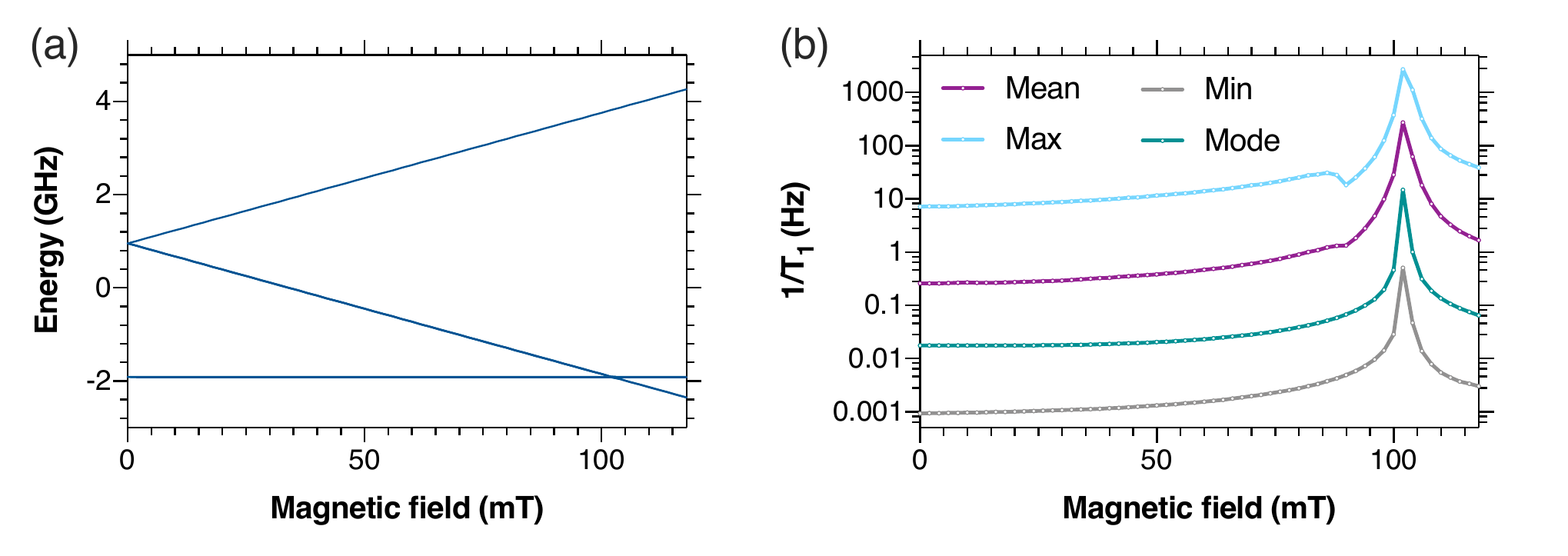}
	\caption{ Spin relaxation in $^{13}$C nuclear spin environment. (a) Energy level structure of NV-$^{13}$C system as a function of external magnetic field.  (b) Magnetic field dependence of the spin relaxation rate ($1/T_1$). Light blue, plum, teal, and gray curves show the largest (Max), average (Mean), most probable (Mode), and lowest (Min) relaxation rates obtained on an ensemble of 50 randomly generated arrangements of 31 $^{13}$C nuclear spin corresponds to 1.07\% abundance.  }
	\label{fig:C13}  
\end{figure}

Next, we investigate NV center-$^{13}$C spin bath systems. The settings for the simulations are similar as for the P1 center environment, except for the concentration of the spin defects, for which we used the natural abundance of $^{13}$C. Due to the fairly simple level structure of the NV center-$^{13}$C nuclear spin system, see Fig.~\ref{fig:C13}(a), the relaxation rate curves shown in Fig.~\ref{fig:C13}(b) exhibit only a single peak at 102~mT that correspond to the GSLAC\citep{IvadyDNP2015}. 

For simplicity, here we use the same, first order cluster approximation as before, i.e.\  cluster systems include the central spin and only one nuclear spin. As the nuclear spins have very long coherence time, the relation $T_2^{13\text{C}} << T_1^{\text{NV}}$, where $T_1^{\text{NV}}$ is solely due to $^{13}$C spins, may not be satisfied. In this case an overestimation of the relaxation rates is expected. Therefore, the results presented in Fig.~\ref{fig:C13}(b) may be considered as an upper bound for the $^{13}$C spin bath induced relaxation.

\begin{figure}[h!]
	\includegraphics[width=0.95\columnwidth]{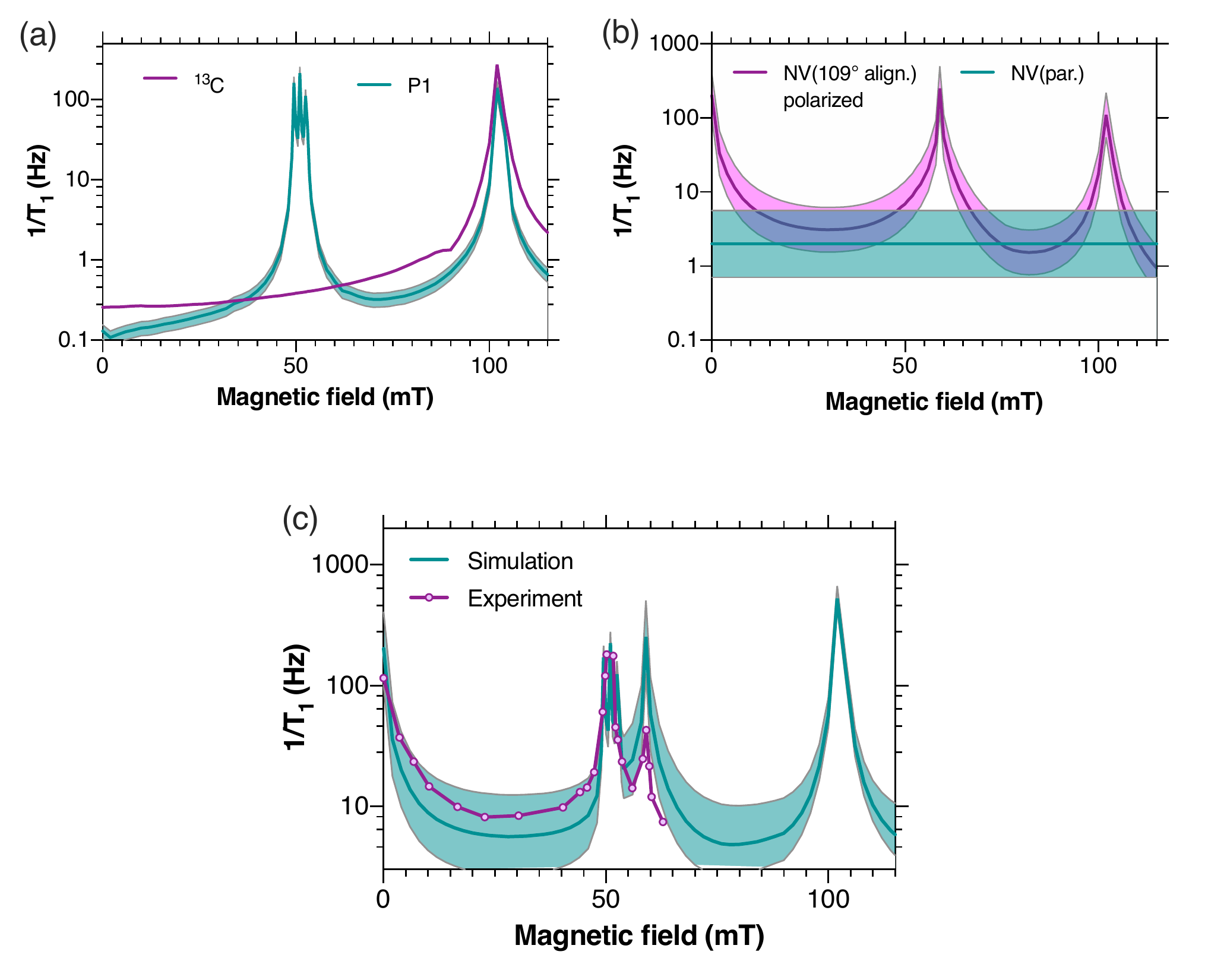}
	\caption{ Comparison between simulation and measurements carried out on sample S2 at 20K in 
	Ref.~[\onlinecite{Jarmola2012}].	
	 (a) and (b) depict theoretical spin relaxation rates for (a) $^{13}$C (plum) and P1 center (teal) spin environments and (b) 109$^{\circ}$ aligned (plum) and parallel (teal) NV center spin environments. Colored areas show estimated uncertainties in the theoretical results due to the error bar of the defect concentrations reported in 
	 Ref.~[\onlinecite{Jarmola2012}].
	 (c) Combined theoretical relaxation rate is compared with the experimental spin relaxation rate reported in
	Ref.~[\onlinecite{Jarmola2012}] 
	for sample S2 at low temperature.}
	\label{fig:errorex}  
\end{figure}

Finally, we combine our theoretical results in order to compare with experimental measurements reported for sample S2 in 
Ref.~[\onlinecite{Jarmola2012}].
The total spin relaxation rate can be given as
\begin{equation}
\frac{1}{T_1^{\text{tot}}} = \frac{1}{T_1^{\text{P1}}}  + \frac{1}{T_1^{\text{cNV-basal}}} + \frac{1}{T_1^{\text{cNV-para}}} + \frac{1}{T_1^{\text{13C}}}    \text{.}
\end{equation}
The theoretical relaxation rate curves with uncertainties deduced from experimental uncertainties in the defect concentrations are depicted in Fig.~\ref{fig:errorex}(a) and (b).  To determine the uncertainties, we use a linear concentration dependence for the relaxation rates\citep{Jarmola2012}. As there is no available data on the strain and magnetic field inhomogeneity nor for the polarization variation of parallel and 109$^{\circ}$ aligned NV centers in the sample modeled here, we make the following assumptions. We assume  1) $\mathcal{O} \!  \left( 0.1 \text{ MHz} \right)$ parallel strain and magnetic field inhomogeneity, 2) $\mathcal{O} \!  \left( 1 \text{ MHz} \right)$ perpendicular strain, and 3) 1\%  variance in the polarization of parallel NV centers. The resultant curves are plotted in Fig.~\ref{fig:errorex}(b). It is apparent from the results that environmental NV centers have a dominant effect on the spin relaxation rate. 

When compared with experiment, we find that the theoretical curve follows the measurements  within error bars over a wide range of the magnetic field considered in the experiment. Higher discrepancy can be seen at $B=59$~mT, where the theoretical curve overestimates the experimental relaxation rate. This can be attributed to the neglect of depolarization of the environmental NV centers. As we have seen in Fig.~\ref{fig:NVstrain}(a), depolarization of the bath reduces relaxation rate. Depolarization of parallel and 109$^{\circ}$ aligned NV centers is expectedly mutual when they couple at $B=59$~mT. Inclusion of this effect can lower the theoretical relaxation rate to the level of experimental measurements.

The numerical results demonstrate that the proposed theoretical method can account for the reported magnetic field dependent spin relaxation patterns induced by P1 centers, NV centers, and $^{13}$C nuclear spins. This is due to the non-approximate description of the pair interactions between the central spin and the environmental spins. Furthermore, numerical simulations validate the approximations introduced by first order cluster approximation in the case of P1 center and NV center spin environments. This makes it possible to obtain quantitative results comparable with experiment.

 \section{Summary}
  \label{sec:disc}
  
In summary this paper describes a microscopic spin bath model for calculating spin relaxation effects in central spin approximation. To this end an extended Lindbladian formalism was introduced to account for spin flip-flops in a many spin system. Validity of the approximation is determined mainly by the relation of environment induced spin flip-flop rates of the central spin and decoherence rate of the spin bath.  The method does not rely on  approximation of the Hamiltonian beyond the central spin approximation. By increasing the order of cluster approximation, errors can be systematically eliminated. 

In the numerical simulations NV center's spin relaxation rate (1/T$_1$) was investigated. P1 center, NV center, and $^{13}$C spin baths are considered at various magnetic fields and strain. The method captures all the known characteristics of the relaxation rate of specific spin bath systems. By taking all the relevant relaxation effects into account, the theoretical spin relaxation rate curve is quantitatively comparable with the measured one over a wide range of magnetic fields.

\section*{Acknowledgments} 

Fruitful discussions with Oscar Bulancea Lindvall are acknowledged. This work was financially supported by the MTA Premium Postdoctoral Research Program. Support from the Knut and Alice Wallenberg Foundation through WBSQD2 project (Grant No. 2018.0071), the Hungarian NKFIH grants No. KKP129866 of
the National Excellence Program of Quantum-coherent materials project, and the NKFIH through the National Quantum Technology Program (Grant No. 2017-1.2.1-NKP-2017-00001) is acknowledged. The numerical simulations were performed on resources provided by the Swedish National Infrastructure for Computing (SNIC) at NSC.

\appendix

\section{Summation approximation}
\label{app:sum_deriv}

Let us consider a closed system $\mathcal{S}$ in central spin arrangement as described in section~\ref{sec:fo_cluster_approx}.  From the general solution of the master equation, one can obtain the reduced density matrix $\varrho_0 \! \left( t \right)$ at a given time $t$ 
\begin{equation}
\varrho_0  \! \left( t \right) =  \text{Tr}_{\text{env}}  \varrho_{\mathcal{S}}   \! \left( t \right)  = \text{Tr}_{\text{env}}  e^{- \frac{i}{\hbar} \left( H_{\mathcal{S}}. - . H_{\mathcal{S}} \right) } \varrho_{\mathcal{S}} \text{,}
\end{equation}
where $\text{Tr}_{\text{env}}$ means trace over all the environmental spin degrees of freedom and 
\begin{equation}
H_{\mathcal{S}} = h_{\text{0} } +  \sum_{i=1}^n \left(  h_{ i } + h_{0 i } \right)  \text{.}
\end{equation}
Let us assume that at time $t$ the system is non-entangled and the density matrix of the many spin system can be written as 
\begin{equation}
\varrho_{\mathcal{S}} = \bigotimes_{i=0}^{n} \varrho_i \text{.}
\end{equation}
Considering an infinitesimal time period $dt$, the reduced density matrix evolves as
\begin{equation}
\varrho_0  \! \left( t + dt \right)  \approx \text{Tr}_{\text{env}} \! \left(  1 - dt \frac{i}{\hbar} \left( H_{\mathcal{S}}. - . H_{\mathcal{S}} \right)  \right)  \varrho_{\mathcal{S}} \text{,}
\end{equation}
from which we obtain
\begin{equation} \label{eq:app1}
d \varrho_0  = - dt \frac{i}{\hbar}  \text{Tr}_{\text{env}}  \left[ h_0 + \sum_i^n \left( h_i + h_{0i} \right) , \bigotimes_{i=0}^{n} \varrho_i \right]  \text{.}
\end{equation} 
By tracing out the unaffected environmental spin degrees of freedom in each terms of the summation, we can rewrite Eq.~(\ref{eq:app1}) as
\begin{equation}
d \varrho_0  = - dt \frac{i}{\hbar} \left(  \left[ h_0, \varrho_0  \right] +  \sum_i \text{Tr}_{i}  \left[  h_{0i} , \varrho_0 \otimes \varrho_i \right] \right)  \text{.}
\end{equation}
The equation above shows that spin flip-flops of the central spin induced by coupling terms $h_{0i}$ are additive for $dt$ time evolution while the spin bath is non-entangled. Note that this argument  can be generalized to partially entangled density matrices, such as 
\begin{equation}
\varrho_{\mathcal{S}}^{i} = \varrho_{0i} \otimes \left( \bigotimes_{j=1, j \neq i}^{n} \varrho_j \right) \text{.}
\end{equation}
for which one gets
\begin{equation}
d \varrho_0  = -dt \frac{i}{\hbar} \left(    \sum_{j \neq i} \text{Tr}_{i,j} \left[ h_0 +  h_i + h_j + h_{0i} + h_{0j} , \varrho_{0i}  \otimes \varrho_j \right]   \right)    \text{.}
\end{equation}

\section{Alternative definition of $da_{0l}$ and $da_{il}$}
\label{app:ail_det}

As mentioned in the main text, complete definition of $da_{0l}$ and $da_{il}$ from the variation of the reduced density matrices $d\varrho_{c_0}^{h\mathcal{E}} $ and $d\varrho_{c_i}^{h\mathcal{E}} $  through Eq.~(\ref{eq:da0l_eq2}) and Eq.~(\ref{eq:dail_eq2}) is not always possible. One can overcome this issue by noticing that $d\varrho_{c_0}^{h\mathcal{E}} $ and $d\varrho_{c_i}^{h\mathcal{E}} $ contains the summed up effect of all the terms in $h_{c_0}$ and $h_{c_i}$. To obtain more equations for  $da_{0l}$ and $da_{il}$  one may split the Hamiltonian and thus the variation of the reduced density matrix into terms like $h_{c_i} = \delta h_{c_i}' + \delta h_{c_i}'' + \delta h_{c_i}''' + ... $  and $d \varrho_{c_i} = \delta \varrho_{c_i}' + \delta \varrho_{c_i}'' + \delta \varrho_{c_i}''' + ... $,where 
\begin{equation}
 \delta \varrho_{c_i}' =  - \frac{i dt}{\hbar} \left[ h_{c_i}',  \varrho_{c_i} \right] \text{.}
\end{equation}
Each $\delta \varrho_{c_i}$  terms may define an independent set of equations similarly to Eq.~(\ref{eq:da0l_eq2}) and Eq.~(\ref{eq:dail_eq2}). This way all ${d_0 \left( d_0 - 1 \right)}/{2}$  transition amplitudes can be determined in all cluster systems in principle.

\bibliographystyle{apsrev4-1}

\begin{thebibliography}{63}%
\makeatletter
\providecommand \@ifxundefined [1]{%
 \@ifx{#1\undefined}
}%
\providecommand \@ifnum [1]{%
 \ifnum #1\expandafter \@firstoftwo
 \else \expandafter \@secondoftwo
 \fi
}%
\providecommand \@ifx [1]{%
 \ifx #1\expandafter \@firstoftwo
 \else \expandafter \@secondoftwo
 \fi
}%
\providecommand \natexlab [1]{#1}%
\providecommand \enquote  [1]{``#1''}%
\providecommand \bibnamefont  [1]{#1}%
\providecommand \bibfnamefont [1]{#1}%
\providecommand \citenamefont [1]{#1}%
\providecommand \href@noop [0]{\@secondoftwo}%
\providecommand \href [0]{\begingroup \@sanitize@url \@href}%
\providecommand \@href[1]{\@@startlink{#1}\@@href}%
\providecommand \@@href[1]{\endgroup#1\@@endlink}%
\providecommand \@sanitize@url [0]{\catcode `\\12\catcode `\$12\catcode
  `\&12\catcode `\#12\catcode `\^12\catcode `\_12\catcode `\%12\relax}%
\providecommand \@@startlink[1]{}%
\providecommand \@@endlink[0]{}%
\providecommand \url  [0]{\begingroup\@sanitize@url \@url }%
\providecommand \@url [1]{\endgroup\@href {#1}{\urlprefix }}%
\providecommand \urlprefix  [0]{URL }%
\providecommand \Eprint [0]{\href }%
\providecommand \doibase [0]{http://dx.doi.org/}%
\providecommand \selectlanguage [0]{\@gobble}%
\providecommand \bibinfo  [0]{\@secondoftwo}%
\providecommand \bibfield  [0]{\@secondoftwo}%
\providecommand \translation [1]{[#1]}%
\providecommand \BibitemOpen [0]{}%
\providecommand \bibitemStop [0]{}%
\providecommand \bibitemNoStop [0]{.\EOS\space}%
\providecommand \EOS [0]{\spacefactor3000\relax}%
\providecommand \BibitemShut  [1]{\csname bibitem#1\endcsname}%
\let\auto@bib@innerbib\@empty
\bibitem [{\citenamefont {du~Preez}(1965)}]{duPreez:1965}%
  \BibitemOpen
  \bibfield  {author} {\bibinfo {author} {\bibfnamefont {L.}~\bibnamefont
  {du~Preez}},\ }\href@noop {} {Ph.D. thesis},\ \bibinfo  {school} {University
  of Witwatersrand} (\bibinfo {year} {1965})\BibitemShut {NoStop}%
\bibitem [{\citenamefont {Wrachtrup}\ and\ \citenamefont
  {Jelezko}(2006)}]{Wrachtrup:JPCM2006}%
  \BibitemOpen
  \bibfield  {author} {\bibinfo {author} {\bibfnamefont {J.}~\bibnamefont
  {Wrachtrup}}\ and\ \bibinfo {author} {\bibfnamefont {F.}~\bibnamefont
  {Jelezko}},\ }\href
  {http://www.iop.org/EJ/article/0953-8984/18/21/S08/cm6_21_S08.pdf} {\bibfield
   {journal} {\bibinfo  {journal} {Journal of Physics-Condensed Matter}\
  }\textbf {\bibinfo {volume} {18}},\ \bibinfo {pages} {S807} (\bibinfo {year}
  {2006})}\BibitemShut {NoStop}%
\bibitem [{\citenamefont {Maze}\ \emph {et~al.}(2011)\citenamefont {Maze},
  \citenamefont {Gali}, \citenamefont {Togan}, \citenamefont {Chu},
  \citenamefont {Trifonov}, \citenamefont {Kaxiras},\ and\ \citenamefont
  {Lukin}}]{Maze2011}%
  \BibitemOpen
  \bibfield  {author} {\bibinfo {author} {\bibfnamefont {J.~R.}\ \bibnamefont
  {Maze}}, \bibinfo {author} {\bibfnamefont {A.}~\bibnamefont {Gali}}, \bibinfo
  {author} {\bibfnamefont {E.}~\bibnamefont {Togan}}, \bibinfo {author}
  {\bibfnamefont {Y.}~\bibnamefont {Chu}}, \bibinfo {author} {\bibfnamefont
  {A.}~\bibnamefont {Trifonov}}, \bibinfo {author} {\bibfnamefont
  {E.}~\bibnamefont {Kaxiras}}, \ and\ \bibinfo {author} {\bibfnamefont
  {M.~D.}\ \bibnamefont {Lukin}},\ }\href
  {http://stacks.iop.org/1367-2630/13/i=2/a=025025} {\bibfield  {journal}
  {\bibinfo  {journal} {New Journal of Physics}\ }\textbf {\bibinfo {volume}
  {13}},\ \bibinfo {pages} {025025} (\bibinfo {year} {2011})}\BibitemShut
  {NoStop}%
\bibitem [{\citenamefont {Doherty}\ \emph {et~al.}(2013)\citenamefont
  {Doherty}, \citenamefont {Manson}, \citenamefont {Delaney}, \citenamefont
  {Jelezko}, \citenamefont {Wrachtrup},\ and\ \citenamefont
  {Hollenberg}}]{DOHERTY20131}%
  \BibitemOpen
  \bibfield  {author} {\bibinfo {author} {\bibfnamefont {M.~W.}\ \bibnamefont
  {Doherty}}, \bibinfo {author} {\bibfnamefont {N.~B.}\ \bibnamefont {Manson}},
  \bibinfo {author} {\bibfnamefont {P.}~\bibnamefont {Delaney}}, \bibinfo
  {author} {\bibfnamefont {F.}~\bibnamefont {Jelezko}}, \bibinfo {author}
  {\bibfnamefont {J.}~\bibnamefont {Wrachtrup}}, \ and\ \bibinfo {author}
  {\bibfnamefont {L.~C.}\ \bibnamefont {Hollenberg}},\ }\href {\doibase
  https://doi.org/10.1016/j.physrep.2013.02.001} {\bibfield  {journal}
  {\bibinfo  {journal} {Physics Reports}\ }\textbf {\bibinfo {volume} {528}},\
  \bibinfo {pages} {1 } (\bibinfo {year} {2013})}\BibitemShut {NoStop}%
\bibitem [{\citenamefont {Gali}(2019)}]{GaliReview2019}%
  \BibitemOpen
  \bibfield  {author} {\bibinfo {author} {\bibfnamefont {A.}~\bibnamefont
  {Gali}},\ }\href {\doibase https://doi:10.1515/nanoph-2019-0154} {\bibfield
  {journal} {\bibinfo  {journal} {Nanophotonics}\ }\textbf {\bibinfo {volume}
  {8}},\ \bibinfo {pages} {1907 } (\bibinfo {year} {2019})}\BibitemShut
  {NoStop}%
\bibitem [{\citenamefont {Gruber}\ \emph {et~al.}(1997)\citenamefont {Gruber},
  \citenamefont {Drabenstedt}, \citenamefont {Tietz}, \citenamefont {Fleury},
  \citenamefont {Wrachtrup},\ and\ \citenamefont
  {Borczyskowski}}]{Gruber:Science1997}%
  \BibitemOpen
  \bibfield  {author} {\bibinfo {author} {\bibfnamefont {A.}~\bibnamefont
  {Gruber}}, \bibinfo {author} {\bibfnamefont {A.}~\bibnamefont {Drabenstedt}},
  \bibinfo {author} {\bibfnamefont {C.}~\bibnamefont {Tietz}}, \bibinfo
  {author} {\bibfnamefont {L.}~\bibnamefont {Fleury}}, \bibinfo {author}
  {\bibfnamefont {J.}~\bibnamefont {Wrachtrup}}, \ and\ \bibinfo {author}
  {\bibfnamefont {C.~v.}\ \bibnamefont {Borczyskowski}},\ }\href {\doibase
  10.1126/science.276.5321.2012} {\bibfield  {journal} {\bibinfo  {journal}
  {Science}\ }\textbf {\bibinfo {volume} {276}},\ \bibinfo {pages} {2012}
  (\bibinfo {year} {1997})}\BibitemShut {NoStop}%
\bibitem [{\citenamefont {Jelezko}\ \emph {et~al.}(2004)\citenamefont
  {Jelezko}, \citenamefont {Gaebel}, \citenamefont {Popa}, \citenamefont
  {Gruber},\ and\ \citenamefont {Wrachtrup}}]{Jelezko:PRL200492}%
  \BibitemOpen
  \bibfield  {author} {\bibinfo {author} {\bibfnamefont {F.}~\bibnamefont
  {Jelezko}}, \bibinfo {author} {\bibfnamefont {T.}~\bibnamefont {Gaebel}},
  \bibinfo {author} {\bibfnamefont {I.}~\bibnamefont {Popa}}, \bibinfo {author}
  {\bibfnamefont {A.}~\bibnamefont {Gruber}}, \ and\ \bibinfo {author}
  {\bibfnamefont {J.}~\bibnamefont {Wrachtrup}},\ }\href
  {http://link.aps.org/abstract/PRL/v92/e076401} {\bibfield  {journal}
  {\bibinfo  {journal} {Physical Review Letters}\ }\textbf {\bibinfo {volume}
  {92}},\ \bibinfo {pages} {076401} (\bibinfo {year} {2004})}\BibitemShut
  {NoStop}%
\bibitem [{\citenamefont {Siyushev}\ \emph {et~al.}(2019)\citenamefont
  {Siyushev}, \citenamefont {Nesladek}, \citenamefont {Bourgeois},
  \citenamefont {Gulka}, \citenamefont {Hruby}, \citenamefont {Yamamoto},
  \citenamefont {Trupke}, \citenamefont {Teraji}, \citenamefont {Isoya},\ and\
  \citenamefont {Jelezko}}]{Siyushev728}%
  \BibitemOpen
  \bibfield  {author} {\bibinfo {author} {\bibfnamefont {P.}~\bibnamefont
  {Siyushev}}, \bibinfo {author} {\bibfnamefont {M.}~\bibnamefont {Nesladek}},
  \bibinfo {author} {\bibfnamefont {E.}~\bibnamefont {Bourgeois}}, \bibinfo
  {author} {\bibfnamefont {M.}~\bibnamefont {Gulka}}, \bibinfo {author}
  {\bibfnamefont {J.}~\bibnamefont {Hruby}}, \bibinfo {author} {\bibfnamefont
  {T.}~\bibnamefont {Yamamoto}}, \bibinfo {author} {\bibfnamefont
  {M.}~\bibnamefont {Trupke}}, \bibinfo {author} {\bibfnamefont
  {T.}~\bibnamefont {Teraji}}, \bibinfo {author} {\bibfnamefont
  {J.}~\bibnamefont {Isoya}}, \ and\ \bibinfo {author} {\bibfnamefont
  {F.}~\bibnamefont {Jelezko}},\ }\href {\doibase 10.1126/science.aav2789}
  {\bibfield  {journal} {\bibinfo  {journal} {Science}\ }\textbf {\bibinfo
  {volume} {363}},\ \bibinfo {pages} {728} (\bibinfo {year}
  {2019})}\BibitemShut {NoStop}%
\bibitem [{\citenamefont {Childress}\ \emph {et~al.}(2006)\citenamefont
  {Childress}, \citenamefont {Gurudev~Dutt}, \citenamefont {Taylor},
  \citenamefont {Zibrov}, \citenamefont {Jelezko}, \citenamefont {Wrachtrup},
  \citenamefont {Hemmer},\ and\ \citenamefont {Lukin}}]{Childress:Science2006}%
  \BibitemOpen
  \bibfield  {author} {\bibinfo {author} {\bibfnamefont {L.}~\bibnamefont
  {Childress}}, \bibinfo {author} {\bibfnamefont {M.~V.}\ \bibnamefont
  {Gurudev~Dutt}}, \bibinfo {author} {\bibfnamefont {J.~M.}\ \bibnamefont
  {Taylor}}, \bibinfo {author} {\bibfnamefont {A.~S.}\ \bibnamefont {Zibrov}},
  \bibinfo {author} {\bibfnamefont {F.}~\bibnamefont {Jelezko}}, \bibinfo
  {author} {\bibfnamefont {J.}~\bibnamefont {Wrachtrup}}, \bibinfo {author}
  {\bibfnamefont {P.~R.}\ \bibnamefont {Hemmer}}, \ and\ \bibinfo {author}
  {\bibfnamefont {M.~D.}\ \bibnamefont {Lukin}},\ }\href@noop {} {\bibfield
  {journal} {\bibinfo  {journal} {Science}\ }\textbf {\bibinfo {volume}
  {314}},\ \bibinfo {pages} {281} (\bibinfo {year} {2006})}\BibitemShut
  {NoStop}%
\bibitem [{\citenamefont {Balasubramanian}\ \emph {et~al.}(2009)\citenamefont
  {Balasubramanian}, \citenamefont {Neumann}, \citenamefont {Twitchen},
  \citenamefont {Markham}, \citenamefont {Kolesov}, \citenamefont {Mizuochi},
  \citenamefont {Isoya}, \citenamefont {Achard}, \citenamefont {Beck},
  \citenamefont {Tissler}, \citenamefont {Jacques}, \citenamefont {Hemmer},
  \citenamefont {Jelezko},\ and\ \citenamefont
  {Wrachtrup}}]{Balasubramanian:NatMat2009}%
  \BibitemOpen
  \bibfield  {author} {\bibinfo {author} {\bibfnamefont {G.}~\bibnamefont
  {Balasubramanian}}, \bibinfo {author} {\bibfnamefont {P.}~\bibnamefont
  {Neumann}}, \bibinfo {author} {\bibfnamefont {D.}~\bibnamefont {Twitchen}},
  \bibinfo {author} {\bibfnamefont {M.}~\bibnamefont {Markham}}, \bibinfo
  {author} {\bibfnamefont {R.}~\bibnamefont {Kolesov}}, \bibinfo {author}
  {\bibfnamefont {N.}~\bibnamefont {Mizuochi}}, \bibinfo {author}
  {\bibfnamefont {J.}~\bibnamefont {Isoya}}, \bibinfo {author} {\bibfnamefont
  {J.}~\bibnamefont {Achard}}, \bibinfo {author} {\bibfnamefont
  {J.}~\bibnamefont {Beck}}, \bibinfo {author} {\bibfnamefont {J.}~\bibnamefont
  {Tissler}}, \bibinfo {author} {\bibfnamefont {V.}~\bibnamefont {Jacques}},
  \bibinfo {author} {\bibfnamefont {P.~R.}\ \bibnamefont {Hemmer}}, \bibinfo
  {author} {\bibfnamefont {F.}~\bibnamefont {Jelezko}}, \ and\ \bibinfo
  {author} {\bibfnamefont {J.}~\bibnamefont {Wrachtrup}},\ }\href
  {http://dx.doi.org/10.1038/nmat2420} {\bibfield  {journal} {\bibinfo
  {journal} {Nat. Mater.}\ }\textbf {\bibinfo {volume} {8}},\ \bibinfo {pages}
  {383} (\bibinfo {year} {2009})}\BibitemShut {NoStop}%
\bibitem [{\citenamefont {Toyli}\ \emph {et~al.}(2012)\citenamefont {Toyli},
  \citenamefont {Christle}, \citenamefont {Alkauskas}, \citenamefont {Buckley},
  \citenamefont {Van~de Walle},\ and\ \citenamefont
  {Awschalom}}]{ToyliPhysRevX2012}%
  \BibitemOpen
  \bibfield  {author} {\bibinfo {author} {\bibfnamefont {D.~M.}\ \bibnamefont
  {Toyli}}, \bibinfo {author} {\bibfnamefont {D.~J.}\ \bibnamefont {Christle}},
  \bibinfo {author} {\bibfnamefont {A.}~\bibnamefont {Alkauskas}}, \bibinfo
  {author} {\bibfnamefont {B.~B.}\ \bibnamefont {Buckley}}, \bibinfo {author}
  {\bibfnamefont {C.~G.}\ \bibnamefont {Van~de Walle}}, \ and\ \bibinfo
  {author} {\bibfnamefont {D.~D.}\ \bibnamefont {Awschalom}},\ }\href {\doibase
  10.1103/PhysRevX.2.031001} {\bibfield  {journal} {\bibinfo  {journal} {Phys.
  Rev. X}\ }\textbf {\bibinfo {volume} {2}},\ \bibinfo {pages} {031001}
  (\bibinfo {year} {2012})}\BibitemShut {NoStop}%
\bibitem [{\citenamefont {Maze}\ \emph
  {et~al.}(2008{\natexlab{a}})\citenamefont {Maze}, \citenamefont {Stanwix},
  \citenamefont {Hodges}, \citenamefont {Hong}, \citenamefont {Taylor},
  \citenamefont {Cappellaro}, \citenamefont {Jiang}, \citenamefont {Dutt},
  \citenamefont {Togan}, \citenamefont {Zibrov}, \citenamefont {Yacoby},
  \citenamefont {Walsworth},\ and\ \citenamefont {Lukin}}]{Maze:Nature2008}%
  \BibitemOpen
  \bibfield  {author} {\bibinfo {author} {\bibfnamefont {J.~R.}\ \bibnamefont
  {Maze}}, \bibinfo {author} {\bibfnamefont {P.~L.}\ \bibnamefont {Stanwix}},
  \bibinfo {author} {\bibfnamefont {J.~S.}\ \bibnamefont {Hodges}}, \bibinfo
  {author} {\bibfnamefont {S.}~\bibnamefont {Hong}}, \bibinfo {author}
  {\bibfnamefont {J.~M.}\ \bibnamefont {Taylor}}, \bibinfo {author}
  {\bibfnamefont {P.}~\bibnamefont {Cappellaro}}, \bibinfo {author}
  {\bibfnamefont {L.}~\bibnamefont {Jiang}}, \bibinfo {author} {\bibfnamefont
  {M.~V.~G.}\ \bibnamefont {Dutt}}, \bibinfo {author} {\bibfnamefont
  {E.}~\bibnamefont {Togan}}, \bibinfo {author} {\bibfnamefont {A.~S.}\
  \bibnamefont {Zibrov}}, \bibinfo {author} {\bibfnamefont {A.}~\bibnamefont
  {Yacoby}}, \bibinfo {author} {\bibfnamefont {R.~L.}\ \bibnamefont
  {Walsworth}}, \ and\ \bibinfo {author} {\bibfnamefont {M.~D.}\ \bibnamefont
  {Lukin}},\ }\href
  {http://www.nature.com/nature/journal/v455/n7213/abs/nature07279.html}
  {\bibfield  {journal} {\bibinfo  {journal} {Nature}\ }\textbf {\bibinfo
  {volume} {455}},\ \bibinfo {pages} {644} (\bibinfo {year}
  {2008}{\natexlab{a}})}\BibitemShut {NoStop}%
\bibitem [{\citenamefont {Dolde}\ \emph {et~al.}(2011)\citenamefont {Dolde},
  \citenamefont {Fedder}, \citenamefont {Doherty}, \citenamefont {N{\"o}bauer},
  \citenamefont {Rempp}, \citenamefont {Balasubramanian}, \citenamefont {Wolf},
  \citenamefont {Reinhard}, \citenamefont {Hollenberg}, \citenamefont
  {Jelezko},\ and\ \citenamefont {Wrachtrup}}]{Dolde2011}%
  \BibitemOpen
  \bibfield  {author} {\bibinfo {author} {\bibfnamefont {F.}~\bibnamefont
  {Dolde}}, \bibinfo {author} {\bibfnamefont {H.}~\bibnamefont {Fedder}},
  \bibinfo {author} {\bibfnamefont {M.~W.}\ \bibnamefont {Doherty}}, \bibinfo
  {author} {\bibfnamefont {T.}~\bibnamefont {N{\"o}bauer}}, \bibinfo {author}
  {\bibfnamefont {F.}~\bibnamefont {Rempp}}, \bibinfo {author} {\bibfnamefont
  {G.}~\bibnamefont {Balasubramanian}}, \bibinfo {author} {\bibfnamefont
  {T.}~\bibnamefont {Wolf}}, \bibinfo {author} {\bibfnamefont {F.}~\bibnamefont
  {Reinhard}}, \bibinfo {author} {\bibfnamefont {L.~C.~L.}\ \bibnamefont
  {Hollenberg}}, \bibinfo {author} {\bibfnamefont {F.}~\bibnamefont {Jelezko}},
  \ and\ \bibinfo {author} {\bibfnamefont {J.}~\bibnamefont {Wrachtrup}},\
  }\href {https://doi.org/10.1038/nphys1969} {\bibfield  {journal} {\bibinfo
  {journal} {Nature Physics}\ }\textbf {\bibinfo {volume} {7}},\ \bibinfo
  {pages} {459} (\bibinfo {year} {2011})}\BibitemShut {NoStop}%
\bibitem [{\citenamefont {Kucsko}\ \emph {et~al.}(2013)\citenamefont {Kucsko},
  \citenamefont {Maurer}, \citenamefont {Yao}, \citenamefont {Kubo},
  \citenamefont {Noh}, \citenamefont {Lo}, \citenamefont {Park},\ and\
  \citenamefont {Lukin}}]{Kucsko2013}%
  \BibitemOpen
  \bibfield  {author} {\bibinfo {author} {\bibfnamefont {G.}~\bibnamefont
  {Kucsko}}, \bibinfo {author} {\bibfnamefont {P.~C.}\ \bibnamefont {Maurer}},
  \bibinfo {author} {\bibfnamefont {N.~Y.}\ \bibnamefont {Yao}}, \bibinfo
  {author} {\bibfnamefont {M.}~\bibnamefont {Kubo}}, \bibinfo {author}
  {\bibfnamefont {H.~J.}\ \bibnamefont {Noh}}, \bibinfo {author} {\bibfnamefont
  {P.~K.}\ \bibnamefont {Lo}}, \bibinfo {author} {\bibfnamefont
  {H.}~\bibnamefont {Park}}, \ and\ \bibinfo {author} {\bibfnamefont {M.~D.}\
  \bibnamefont {Lukin}},\ }\href {\doibase 10.1038/nature12373} {\bibfield
  {journal} {\bibinfo  {journal} {Nature}\ }\textbf {\bibinfo {volume} {500}},\
  \bibinfo {pages} {54} (\bibinfo {year} {2013})}\BibitemShut {NoStop}%
\bibitem [{\citenamefont {Teissier}\ \emph {et~al.}(2014)\citenamefont
  {Teissier}, \citenamefont {Barfuss}, \citenamefont {Appel}, \citenamefont
  {Neu},\ and\ \citenamefont {Maletinsky}}]{Teissier2014}%
  \BibitemOpen
  \bibfield  {author} {\bibinfo {author} {\bibfnamefont {J.}~\bibnamefont
  {Teissier}}, \bibinfo {author} {\bibfnamefont {A.}~\bibnamefont {Barfuss}},
  \bibinfo {author} {\bibfnamefont {P.}~\bibnamefont {Appel}}, \bibinfo
  {author} {\bibfnamefont {E.}~\bibnamefont {Neu}}, \ and\ \bibinfo {author}
  {\bibfnamefont {P.}~\bibnamefont {Maletinsky}},\ }\href {\doibase
  10.1103/PhysRevLett.113.020503} {\bibfield  {journal} {\bibinfo  {journal}
  {Phys. Rev. Lett.}\ }\textbf {\bibinfo {volume} {113}},\ \bibinfo {pages}
  {020503} (\bibinfo {year} {2014})}\BibitemShut {NoStop}%
\bibitem [{\citenamefont {Weber}\ \emph {et~al.}(2010)\citenamefont {Weber},
  \citenamefont {Koehl}, \citenamefont {Varley}, \citenamefont {Janotti},
  \citenamefont {Buckley}, \citenamefont {Van~de Walle},\ and\ \citenamefont
  {Awschalom}}]{Weber10}%
  \BibitemOpen
  \bibfield  {author} {\bibinfo {author} {\bibfnamefont {J.~R.}\ \bibnamefont
  {Weber}}, \bibinfo {author} {\bibfnamefont {W.~F.}\ \bibnamefont {Koehl}},
  \bibinfo {author} {\bibfnamefont {J.~B.}\ \bibnamefont {Varley}}, \bibinfo
  {author} {\bibfnamefont {A.}~\bibnamefont {Janotti}}, \bibinfo {author}
  {\bibfnamefont {B.~B.}\ \bibnamefont {Buckley}}, \bibinfo {author}
  {\bibfnamefont {C.~G.}\ \bibnamefont {Van~de Walle}}, \ and\ \bibinfo
  {author} {\bibfnamefont {D.~D.}\ \bibnamefont {Awschalom}},\ }\href {\doibase
  10.1073/pnas.1003052107} {\bibfield  {journal} {\bibinfo  {journal} {PNAS}\
  }\textbf {\bibinfo {volume} {107}},\ \bibinfo {pages} {8513} (\bibinfo {year}
  {2010})}\BibitemShut {NoStop}%
\bibitem [{\citenamefont {Awschalom}\ \emph {et~al.}(2013)\citenamefont
  {Awschalom}, \citenamefont {Bassett}, \citenamefont {Dzurak}, \citenamefont
  {Hu},\ and\ \citenamefont {Petta}}]{Awschalom2013}%
  \BibitemOpen
  \bibfield  {author} {\bibinfo {author} {\bibfnamefont {D.~D.}\ \bibnamefont
  {Awschalom}}, \bibinfo {author} {\bibfnamefont {L.~C.}\ \bibnamefont
  {Bassett}}, \bibinfo {author} {\bibfnamefont {A.~S.}\ \bibnamefont {Dzurak}},
  \bibinfo {author} {\bibfnamefont {E.~L.}\ \bibnamefont {Hu}}, \ and\ \bibinfo
  {author} {\bibfnamefont {J.~R.}\ \bibnamefont {Petta}},\ }\href {\doibase
  10.1126/science.1231364} {\bibfield  {journal} {\bibinfo  {journal}
  {Science}\ }\textbf {\bibinfo {volume} {339}},\ \bibinfo {pages} {1174}
  (\bibinfo {year} {2013})}\BibitemShut {NoStop}%
\bibitem [{\citenamefont {Koehl}\ \emph {et~al.}(2011)\citenamefont {Koehl},
  \citenamefont {Buckley}, \citenamefont {Heremans}, \citenamefont {Calusine},\
  and\ \citenamefont {Awschalom}}]{Koehl11}%
  \BibitemOpen
  \bibfield  {author} {\bibinfo {author} {\bibfnamefont {W.~F.}\ \bibnamefont
  {Koehl}}, \bibinfo {author} {\bibfnamefont {B.~B.}\ \bibnamefont {Buckley}},
  \bibinfo {author} {\bibfnamefont {F.~J.}\ \bibnamefont {Heremans}}, \bibinfo
  {author} {\bibfnamefont {G.}~\bibnamefont {Calusine}}, \ and\ \bibinfo
  {author} {\bibfnamefont {D.~D.}\ \bibnamefont {Awschalom}},\ }\href@noop {}
  {\bibfield  {journal} {\bibinfo  {journal} {Nature}\ }\textbf {\bibinfo
  {volume} {479}},\ \bibinfo {pages} {84} (\bibinfo {year} {2011})}\BibitemShut
  {NoStop}%
\bibitem [{\citenamefont {Widmann}\ \emph {et~al.}(2015)\citenamefont
  {Widmann}, \citenamefont {Lee}, \citenamefont {Rendler}, \citenamefont {Son},
  \citenamefont {Fedder}, \citenamefont {Paik}, \citenamefont {Yang},
  \citenamefont {Zhao}, \citenamefont {Yang}, \citenamefont {Booker},
  \citenamefont {Denisenko}, \citenamefont {Jamali}, \citenamefont
  {Momenzadeh}, \citenamefont {Gerhardt}, \citenamefont {Ohshima},
  \citenamefont {Gali}, \citenamefont {Janz{\'e}n},\ and\ \citenamefont
  {Wrachtrup}}]{Widmann2015}%
  \BibitemOpen
  \bibfield  {author} {\bibinfo {author} {\bibfnamefont {M.}~\bibnamefont
  {Widmann}}, \bibinfo {author} {\bibfnamefont {S.-Y.}\ \bibnamefont {Lee}},
  \bibinfo {author} {\bibfnamefont {T.}~\bibnamefont {Rendler}}, \bibinfo
  {author} {\bibfnamefont {N.~T.}\ \bibnamefont {Son}}, \bibinfo {author}
  {\bibfnamefont {H.}~\bibnamefont {Fedder}}, \bibinfo {author} {\bibfnamefont
  {S.}~\bibnamefont {Paik}}, \bibinfo {author} {\bibfnamefont {L.-P.}\
  \bibnamefont {Yang}}, \bibinfo {author} {\bibfnamefont {N.}~\bibnamefont
  {Zhao}}, \bibinfo {author} {\bibfnamefont {S.}~\bibnamefont {Yang}}, \bibinfo
  {author} {\bibfnamefont {I.}~\bibnamefont {Booker}}, \bibinfo {author}
  {\bibfnamefont {A.}~\bibnamefont {Denisenko}}, \bibinfo {author}
  {\bibfnamefont {M.}~\bibnamefont {Jamali}}, \bibinfo {author} {\bibfnamefont
  {S.~A.}\ \bibnamefont {Momenzadeh}}, \bibinfo {author} {\bibfnamefont
  {I.}~\bibnamefont {Gerhardt}}, \bibinfo {author} {\bibfnamefont
  {T.}~\bibnamefont {Ohshima}}, \bibinfo {author} {\bibfnamefont
  {A.}~\bibnamefont {Gali}}, \bibinfo {author} {\bibfnamefont {E.}~\bibnamefont
  {Janz{\'e}n}}, \ and\ \bibinfo {author} {\bibfnamefont {J.}~\bibnamefont
  {Wrachtrup}},\ }\href {\doibase 10.1038/nmat4145} {\bibfield  {journal}
  {\bibinfo  {journal} {Nature Materials}\ }\textbf {\bibinfo {volume} {14}},\
  \bibinfo {pages} {164} (\bibinfo {year} {2015})}\BibitemShut {NoStop}%
\bibitem [{\citenamefont {Rose}\ \emph {et~al.}(2018)\citenamefont {Rose},
  \citenamefont {Huang}, \citenamefont {Zhang}, \citenamefont {Stevenson},
  \citenamefont {Tyryshkin}, \citenamefont {Sangtawesin}, \citenamefont
  {Srinivasan}, \citenamefont {Loudin}, \citenamefont {Markham}, \citenamefont
  {Edmonds}, \citenamefont {Twitchen}, \citenamefont {Lyon},\ and\
  \citenamefont {de~Leon}}]{Rose2018}%
  \BibitemOpen
  \bibfield  {author} {\bibinfo {author} {\bibfnamefont {B.~C.}\ \bibnamefont
  {Rose}}, \bibinfo {author} {\bibfnamefont {D.}~\bibnamefont {Huang}},
  \bibinfo {author} {\bibfnamefont {Z.-H.}\ \bibnamefont {Zhang}}, \bibinfo
  {author} {\bibfnamefont {P.}~\bibnamefont {Stevenson}}, \bibinfo {author}
  {\bibfnamefont {A.~M.}\ \bibnamefont {Tyryshkin}}, \bibinfo {author}
  {\bibfnamefont {S.}~\bibnamefont {Sangtawesin}}, \bibinfo {author}
  {\bibfnamefont {S.}~\bibnamefont {Srinivasan}}, \bibinfo {author}
  {\bibfnamefont {L.}~\bibnamefont {Loudin}}, \bibinfo {author} {\bibfnamefont
  {M.~L.}\ \bibnamefont {Markham}}, \bibinfo {author} {\bibfnamefont {A.~M.}\
  \bibnamefont {Edmonds}}, \bibinfo {author} {\bibfnamefont {D.~J.}\
  \bibnamefont {Twitchen}}, \bibinfo {author} {\bibfnamefont {S.~A.}\
  \bibnamefont {Lyon}}, \ and\ \bibinfo {author} {\bibfnamefont {N.~P.}\
  \bibnamefont {de~Leon}},\ }\href {\doibase 10.1126/science.aao0290}
  {\bibfield  {journal} {\bibinfo  {journal} {Science}\ }\textbf {\bibinfo
  {volume} {361}},\ \bibinfo {pages} {60} (\bibinfo {year} {2018})}\BibitemShut
  {NoStop}%
\bibitem [{\citenamefont {Broadway}\ \emph {et~al.}(2018)\citenamefont
  {Broadway}, \citenamefont {Tetienne}, \citenamefont {Stacey}, \citenamefont
  {Wood}, \citenamefont {Simpson}, \citenamefont {Hall},\ and\ \citenamefont
  {Hollenberg}}]{Broadway2018}%
  \BibitemOpen
  \bibfield  {author} {\bibinfo {author} {\bibfnamefont {D.~A.}\ \bibnamefont
  {Broadway}}, \bibinfo {author} {\bibfnamefont {J.-P.}\ \bibnamefont
  {Tetienne}}, \bibinfo {author} {\bibfnamefont {A.}~\bibnamefont {Stacey}},
  \bibinfo {author} {\bibfnamefont {J.~D.~A.}\ \bibnamefont {Wood}}, \bibinfo
  {author} {\bibfnamefont {D.~A.}\ \bibnamefont {Simpson}}, \bibinfo {author}
  {\bibfnamefont {L.~T.}\ \bibnamefont {Hall}}, \ and\ \bibinfo {author}
  {\bibfnamefont {L.~C.~L.}\ \bibnamefont {Hollenberg}},\ }\href {\doibase
  10.1038/s41467-018-03578-1} {\bibfield  {journal} {\bibinfo  {journal}
  {Nature Communications}\ }\textbf {\bibinfo {volume} {9}},\ \bibinfo {pages}
  {1246} (\bibinfo {year} {2018})}\BibitemShut {NoStop}%
\bibitem [{\citenamefont {Wunderlich}\ \emph {et~al.}(2018)\citenamefont
  {Wunderlich}, \citenamefont {Kohlrautz}, \citenamefont {Abel}, \citenamefont
  {Haase},\ and\ \citenamefont {Meijer}}]{Wunderlich_2018}%
  \BibitemOpen
  \bibfield  {author} {\bibinfo {author} {\bibfnamefont {R.}~\bibnamefont
  {Wunderlich}}, \bibinfo {author} {\bibfnamefont {J.}~\bibnamefont
  {Kohlrautz}}, \bibinfo {author} {\bibfnamefont {B.}~\bibnamefont {Abel}},
  \bibinfo {author} {\bibfnamefont {J.}~\bibnamefont {Haase}}, \ and\ \bibinfo
  {author} {\bibfnamefont {J.}~\bibnamefont {Meijer}},\ }\href {\doibase
  10.1088/1361-648x/aacc32} {\bibfield  {journal} {\bibinfo  {journal} {Journal
  of Physics: Condensed Matter}\ }\textbf {\bibinfo {volume} {30}},\ \bibinfo
  {pages} {305803} (\bibinfo {year} {2018})}\BibitemShut {NoStop}%
\bibitem [{\citenamefont {Witzel}\ \emph {et~al.}(2005)\citenamefont {Witzel},
  \citenamefont {de~Sousa},\ and\ \citenamefont {Das~Sarma}}]{Witzel2005}%
  \BibitemOpen
  \bibfield  {author} {\bibinfo {author} {\bibfnamefont {W.~M.}\ \bibnamefont
  {Witzel}}, \bibinfo {author} {\bibfnamefont {R.}~\bibnamefont {de~Sousa}}, \
  and\ \bibinfo {author} {\bibfnamefont {S.}~\bibnamefont {Das~Sarma}},\ }\href
  {\doibase 10.1103/PhysRevB.72.161306} {\bibfield  {journal} {\bibinfo
  {journal} {Phys. Rev. B}\ }\textbf {\bibinfo {volume} {72}},\ \bibinfo
  {pages} {161306} (\bibinfo {year} {2005})}\BibitemShut {NoStop}%
\bibitem [{\citenamefont {Witzel}\ and\ \citenamefont
  {Sarma}(2006)}]{Witzel:PRB2006}%
  \BibitemOpen
  \bibfield  {author} {\bibinfo {author} {\bibfnamefont {W.~M.}\ \bibnamefont
  {Witzel}}\ and\ \bibinfo {author} {\bibfnamefont {S.~D.}\ \bibnamefont
  {Sarma}},\ }\href@noop {} {\bibfield  {journal} {\bibinfo  {journal} {Phys.
  Rev. B}\ }\textbf {\bibinfo {volume} {74}},\ \bibinfo {eid} {035322}
  (\bibinfo {year} {2006})}\BibitemShut {NoStop}%
\bibitem [{\citenamefont {Saikin}\ \emph {et~al.}(2007)\citenamefont {Saikin},
  \citenamefont {Yao},\ and\ \citenamefont {Sham}}]{ShamLinkedCluster2007}%
  \BibitemOpen
  \bibfield  {author} {\bibinfo {author} {\bibfnamefont {S.~K.}\ \bibnamefont
  {Saikin}}, \bibinfo {author} {\bibfnamefont {W.}~\bibnamefont {Yao}}, \ and\
  \bibinfo {author} {\bibfnamefont {L.~J.}\ \bibnamefont {Sham}},\ }\href
  {\doibase 10.1103/PhysRevB.75.125314} {\bibfield  {journal} {\bibinfo
  {journal} {Phys. Rev. B}\ }\textbf {\bibinfo {volume} {75}},\ \bibinfo
  {pages} {125314} (\bibinfo {year} {2007})}\BibitemShut {NoStop}%
\bibitem [{\citenamefont {Liu}\ \emph {et~al.}(2007)\citenamefont {Liu},
  \citenamefont {Yao},\ and\ \citenamefont {Sham}}]{Liu_2007}%
  \BibitemOpen
  \bibfield  {author} {\bibinfo {author} {\bibfnamefont {R.-B.}\ \bibnamefont
  {Liu}}, \bibinfo {author} {\bibfnamefont {W.}~\bibnamefont {Yao}}, \ and\
  \bibinfo {author} {\bibfnamefont {L.~J.}\ \bibnamefont {Sham}},\ }\href
  {\doibase 10.1088/1367-2630/9/7/226} {\bibfield  {journal} {\bibinfo
  {journal} {New Journal of Physics}\ }\textbf {\bibinfo {volume} {9}},\
  \bibinfo {pages} {226} (\bibinfo {year} {2007})}\BibitemShut {NoStop}%
\bibitem [{\citenamefont {Maze}\ \emph
  {et~al.}(2008{\natexlab{b}})\citenamefont {Maze}, \citenamefont {Taylor},\
  and\ \citenamefont {Lukin}}]{MazePRB2008Decoh}%
  \BibitemOpen
  \bibfield  {author} {\bibinfo {author} {\bibfnamefont {J.~R.}\ \bibnamefont
  {Maze}}, \bibinfo {author} {\bibfnamefont {J.~M.}\ \bibnamefont {Taylor}}, \
  and\ \bibinfo {author} {\bibfnamefont {M.~D.}\ \bibnamefont {Lukin}},\ }\href
  {\doibase 10.1103/PhysRevB.78.094303} {\bibfield  {journal} {\bibinfo
  {journal} {Phys. Rev. B}\ }\textbf {\bibinfo {volume} {78}},\ \bibinfo
  {pages} {094303} (\bibinfo {year} {2008}{\natexlab{b}})}\BibitemShut
  {NoStop}%
\bibitem [{\citenamefont {Taylor}\ \emph {et~al.}(2008)\citenamefont {Taylor},
  \citenamefont {Cappellaro}, \citenamefont {Childress}, \citenamefont {Jiang},
  \citenamefont {Budker}, \citenamefont {Hemmer}, \citenamefont {Yacoby},
  \citenamefont {Walsworth},\ and\ \citenamefont {Lukin}}]{Taylor:NatPhys2008}%
  \BibitemOpen
  \bibfield  {author} {\bibinfo {author} {\bibfnamefont {J.}~\bibnamefont
  {Taylor}}, \bibinfo {author} {\bibfnamefont {P.}~\bibnamefont {Cappellaro}},
  \bibinfo {author} {\bibfnamefont {L.}~\bibnamefont {Childress}}, \bibinfo
  {author} {\bibfnamefont {L.}~\bibnamefont {Jiang}}, \bibinfo {author}
  {\bibfnamefont {D.}~\bibnamefont {Budker}}, \bibinfo {author} {\bibfnamefont
  {P.}~\bibnamefont {Hemmer}}, \bibinfo {author} {\bibfnamefont
  {A.}~\bibnamefont {Yacoby}}, \bibinfo {author} {\bibfnamefont
  {R.}~\bibnamefont {Walsworth}}, \ and\ \bibinfo {author} {\bibfnamefont
  {M.}~\bibnamefont {Lukin}},\ }\href {http://arxiv.org/pdf/0805.1367}
  {\bibfield  {journal} {\bibinfo  {journal} {Nat. Phys.}\ }\textbf {\bibinfo
  {volume} {4}},\ \bibinfo {pages} {810} (\bibinfo {year} {2008})}\BibitemShut
  {NoStop}%
\bibitem [{\citenamefont {Hanson}\ \emph {et~al.}(2008)\citenamefont {Hanson},
  \citenamefont {Dobrovitski}, \citenamefont {Feiguin}, \citenamefont {Gywat},\
  and\ \citenamefont {Awschalom}}]{Hanson:Science2008}%
  \BibitemOpen
  \bibfield  {author} {\bibinfo {author} {\bibfnamefont {R.}~\bibnamefont
  {Hanson}}, \bibinfo {author} {\bibfnamefont {V.~V.}\ \bibnamefont
  {Dobrovitski}}, \bibinfo {author} {\bibfnamefont {A.~E.}\ \bibnamefont
  {Feiguin}}, \bibinfo {author} {\bibfnamefont {O.}~\bibnamefont {Gywat}}, \
  and\ \bibinfo {author} {\bibfnamefont {D.~D.}\ \bibnamefont {Awschalom}},\
  }\href {http://www.sciencemag.org/cgi/content/abstract/320/5874/352}
  {\bibfield  {journal} {\bibinfo  {journal} {Science}\ }\textbf {\bibinfo
  {volume} {320}},\ \bibinfo {pages} {352} (\bibinfo {year}
  {2008})}\BibitemShut {NoStop}%
\bibitem [{\citenamefont {Cywi\ifmmode~\acute{n}\else \'{n}\fi{}ski}\ \emph
  {et~al.}(2009)\citenamefont {Cywi\ifmmode~\acute{n}\else \'{n}\fi{}ski},
  \citenamefont {Witzel},\ and\ \citenamefont {Das~Sarma}}]{Sarma2009}%
  \BibitemOpen
  \bibfield  {author} {\bibinfo {author} {\bibfnamefont {L.}~\bibnamefont
  {Cywi\ifmmode~\acute{n}\else \'{n}\fi{}ski}}, \bibinfo {author}
  {\bibfnamefont {W.~M.}\ \bibnamefont {Witzel}}, \ and\ \bibinfo {author}
  {\bibfnamefont {S.}~\bibnamefont {Das~Sarma}},\ }\href {\doibase
  10.1103/PhysRevB.79.245314} {\bibfield  {journal} {\bibinfo  {journal} {Phys.
  Rev. B}\ }\textbf {\bibinfo {volume} {79}},\ \bibinfo {pages} {245314}
  (\bibinfo {year} {2009})}\BibitemShut {NoStop}%
\bibitem [{\citenamefont {Maze}\ \emph {et~al.}(2012)\citenamefont {Maze},
  \citenamefont {Dr{\'{e}}au}, \citenamefont {Waselowski}, \citenamefont
  {Duarte}, \citenamefont {Roch},\ and\ \citenamefont {Jacques}}]{Maze_2012}%
  \BibitemOpen
  \bibfield  {author} {\bibinfo {author} {\bibfnamefont {J.~R.}\ \bibnamefont
  {Maze}}, \bibinfo {author} {\bibfnamefont {A.}~\bibnamefont {Dr{\'{e}}au}},
  \bibinfo {author} {\bibfnamefont {V.}~\bibnamefont {Waselowski}}, \bibinfo
  {author} {\bibfnamefont {H.}~\bibnamefont {Duarte}}, \bibinfo {author}
  {\bibfnamefont {J.-F.}\ \bibnamefont {Roch}}, \ and\ \bibinfo {author}
  {\bibfnamefont {V.}~\bibnamefont {Jacques}},\ }\href {\doibase
  10.1088/1367-2630/14/10/103041} {\bibfield  {journal} {\bibinfo  {journal}
  {New Journal of Physics}\ }\textbf {\bibinfo {volume} {14}},\ \bibinfo
  {pages} {103041} (\bibinfo {year} {2012})}\BibitemShut {NoStop}%
\bibitem [{\citenamefont {Hall}\ \emph {et~al.}(2014)\citenamefont {Hall},
  \citenamefont {Cole},\ and\ \citenamefont
  {Hollenberg}}]{HollenbergDecoh2014}%
  \BibitemOpen
  \bibfield  {author} {\bibinfo {author} {\bibfnamefont {L.~T.}\ \bibnamefont
  {Hall}}, \bibinfo {author} {\bibfnamefont {J.~H.}\ \bibnamefont {Cole}}, \
  and\ \bibinfo {author} {\bibfnamefont {L.~C.~L.}\ \bibnamefont
  {Hollenberg}},\ }\href {\doibase 10.1103/PhysRevB.90.075201} {\bibfield
  {journal} {\bibinfo  {journal} {Phys. Rev. B}\ }\textbf {\bibinfo {volume}
  {90}},\ \bibinfo {pages} {075201} (\bibinfo {year} {2014})}\BibitemShut
  {NoStop}%
\bibitem [{\citenamefont {Balian}\ \emph {et~al.}(2014)\citenamefont {Balian},
  \citenamefont {Wolfowicz}, \citenamefont {Morton},\ and\ \citenamefont
  {Monteiro}}]{BalianPRB2014}%
  \BibitemOpen
  \bibfield  {author} {\bibinfo {author} {\bibfnamefont {S.~J.}\ \bibnamefont
  {Balian}}, \bibinfo {author} {\bibfnamefont {G.}~\bibnamefont {Wolfowicz}},
  \bibinfo {author} {\bibfnamefont {J.~J.~L.}\ \bibnamefont {Morton}}, \ and\
  \bibinfo {author} {\bibfnamefont {T.~S.}\ \bibnamefont {Monteiro}},\ }\href
  {\doibase 10.1103/PhysRevB.89.045403} {\bibfield  {journal} {\bibinfo
  {journal} {Phys. Rev. B}\ }\textbf {\bibinfo {volume} {89}},\ \bibinfo
  {pages} {045403} (\bibinfo {year} {2014})}\BibitemShut {NoStop}%
\bibitem [{\citenamefont {Wang}\ and\ \citenamefont
  {Takahashi}(2013)}]{Susumu2013}%
  \BibitemOpen
  \bibfield  {author} {\bibinfo {author} {\bibfnamefont {Z.-H.}\ \bibnamefont
  {Wang}}\ and\ \bibinfo {author} {\bibfnamefont {S.}~\bibnamefont
  {Takahashi}},\ }\href {\doibase 10.1103/PhysRevB.87.115122} {\bibfield
  {journal} {\bibinfo  {journal} {Phys. Rev. B}\ }\textbf {\bibinfo {volume}
  {87}},\ \bibinfo {pages} {115122} (\bibinfo {year} {2013})}\BibitemShut
  {NoStop}%
\bibitem [{\citenamefont {Yang}\ and\ \citenamefont
  {Liu}(2008)}]{YangPRB2008CCE}%
  \BibitemOpen
  \bibfield  {author} {\bibinfo {author} {\bibfnamefont {W.}~\bibnamefont
  {Yang}}\ and\ \bibinfo {author} {\bibfnamefont {R.-B.}\ \bibnamefont {Liu}},\
  }\href {\doibase 10.1103/PhysRevB.78.085315} {\bibfield  {journal} {\bibinfo
  {journal} {Phys. Rev. B}\ }\textbf {\bibinfo {volume} {78}},\ \bibinfo
  {pages} {085315} (\bibinfo {year} {2008})}\BibitemShut {NoStop}%
\bibitem [{\citenamefont {Yang}\ and\ \citenamefont
  {Liu}(2009)}]{YangPRB2009CCEEnsamble}%
  \BibitemOpen
  \bibfield  {author} {\bibinfo {author} {\bibfnamefont {W.}~\bibnamefont
  {Yang}}\ and\ \bibinfo {author} {\bibfnamefont {R.-B.}\ \bibnamefont {Liu}},\
  }\href {\doibase 10.1103/PhysRevB.79.115320} {\bibfield  {journal} {\bibinfo
  {journal} {Phys. Rev. B}\ }\textbf {\bibinfo {volume} {79}},\ \bibinfo
  {pages} {115320} (\bibinfo {year} {2009})}\BibitemShut {NoStop}%
\bibitem [{\citenamefont {Zhao}\ \emph {et~al.}(2012)\citenamefont {Zhao},
  \citenamefont {Ho},\ and\ \citenamefont {Liu}}]{RenBaoLiuNVDecoh2012}%
  \BibitemOpen
  \bibfield  {author} {\bibinfo {author} {\bibfnamefont {N.}~\bibnamefont
  {Zhao}}, \bibinfo {author} {\bibfnamefont {S.-W.}\ \bibnamefont {Ho}}, \ and\
  \bibinfo {author} {\bibfnamefont {R.-B.}\ \bibnamefont {Liu}},\ }\href
  {\doibase 10.1103/PhysRevB.85.115303} {\bibfield  {journal} {\bibinfo
  {journal} {Phys. Rev. B}\ }\textbf {\bibinfo {volume} {85}},\ \bibinfo
  {pages} {115303} (\bibinfo {year} {2012})}\BibitemShut {NoStop}%
\bibitem [{\citenamefont {Yang}\ \emph {et~al.}(2014)\citenamefont {Yang},
  \citenamefont {Burk}, \citenamefont {Widmann}, \citenamefont {Lee},
  \citenamefont {Wrachtrup},\ and\ \citenamefont {Zhao}}]{YangPRB2014}%
  \BibitemOpen
  \bibfield  {author} {\bibinfo {author} {\bibfnamefont {L.-P.}\ \bibnamefont
  {Yang}}, \bibinfo {author} {\bibfnamefont {C.}~\bibnamefont {Burk}}, \bibinfo
  {author} {\bibfnamefont {M.}~\bibnamefont {Widmann}}, \bibinfo {author}
  {\bibfnamefont {S.-Y.}\ \bibnamefont {Lee}}, \bibinfo {author} {\bibfnamefont
  {J.}~\bibnamefont {Wrachtrup}}, \ and\ \bibinfo {author} {\bibfnamefont
  {N.}~\bibnamefont {Zhao}},\ }\href {\doibase 10.1103/PhysRevB.90.241203}
  {\bibfield  {journal} {\bibinfo  {journal} {Phys. Rev. B}\ }\textbf {\bibinfo
  {volume} {90}},\ \bibinfo {pages} {241203} (\bibinfo {year}
  {2014})}\BibitemShut {NoStop}%
\bibitem [{\citenamefont {Seo}\ \emph {et~al.}(2016)\citenamefont {Seo},
  \citenamefont {Falk}, \citenamefont {Klimov}, \citenamefont {Miao},
  \citenamefont {Galli},\ and\ \citenamefont {Awschalom}}]{SeoNatureComm2016}%
  \BibitemOpen
  \bibfield  {author} {\bibinfo {author} {\bibfnamefont {H.}~\bibnamefont
  {Seo}}, \bibinfo {author} {\bibfnamefont {A.~L.}\ \bibnamefont {Falk}},
  \bibinfo {author} {\bibfnamefont {P.~V.}\ \bibnamefont {Klimov}}, \bibinfo
  {author} {\bibfnamefont {K.~C.}\ \bibnamefont {Miao}}, \bibinfo {author}
  {\bibfnamefont {G.}~\bibnamefont {Galli}}, \ and\ \bibinfo {author}
  {\bibfnamefont {D.~D.}\ \bibnamefont {Awschalom}},\ }\href
  {https://doi.org/10.1038/ncomms12935} {\bibfield  {journal} {\bibinfo
  {journal} {Nature Communications}\ }\textbf {\bibinfo {volume} {7}},\
  \bibinfo {pages} {12935} (\bibinfo {year} {2016})}\BibitemShut {NoStop}%
\bibitem [{\citenamefont {Doherty}\ \emph {et~al.}(2012)\citenamefont
  {Doherty}, \citenamefont {Dolde}, \citenamefont {Fedder}, \citenamefont
  {Jelezko}, \citenamefont {Wrachtrup}, \citenamefont {Manson},\ and\
  \citenamefont {Hollenberg}}]{Doherty2012}%
  \BibitemOpen
  \bibfield  {author} {\bibinfo {author} {\bibfnamefont {M.~W.}\ \bibnamefont
  {Doherty}}, \bibinfo {author} {\bibfnamefont {F.}~\bibnamefont {Dolde}},
  \bibinfo {author} {\bibfnamefont {H.}~\bibnamefont {Fedder}}, \bibinfo
  {author} {\bibfnamefont {F.}~\bibnamefont {Jelezko}}, \bibinfo {author}
  {\bibfnamefont {J.}~\bibnamefont {Wrachtrup}}, \bibinfo {author}
  {\bibfnamefont {N.~B.}\ \bibnamefont {Manson}}, \ and\ \bibinfo {author}
  {\bibfnamefont {L.~C.~L.}\ \bibnamefont {Hollenberg}},\ }\href {\doibase
  10.1103/PhysRevB.85.205203} {\bibfield  {journal} {\bibinfo  {journal} {Phys.
  Rev. B}\ }\textbf {\bibinfo {volume} {85}},\ \bibinfo {pages} {205203}
  (\bibinfo {year} {2012})}\BibitemShut {NoStop}%
\bibitem [{\citenamefont {Norambuena}\ \emph {et~al.}(2018)\citenamefont
  {Norambuena}, \citenamefont {Mu\~noz}, \citenamefont {Dinani}, \citenamefont
  {Jarmola}, \citenamefont {Maletinsky}, \citenamefont {Budker},\ and\
  \citenamefont {Maze}}]{Norambuena2018}%
  \BibitemOpen
  \bibfield  {author} {\bibinfo {author} {\bibfnamefont {A.}~\bibnamefont
  {Norambuena}}, \bibinfo {author} {\bibfnamefont {E.}~\bibnamefont {Mu\~noz}},
  \bibinfo {author} {\bibfnamefont {H.~T.}\ \bibnamefont {Dinani}}, \bibinfo
  {author} {\bibfnamefont {A.}~\bibnamefont {Jarmola}}, \bibinfo {author}
  {\bibfnamefont {P.}~\bibnamefont {Maletinsky}}, \bibinfo {author}
  {\bibfnamefont {D.}~\bibnamefont {Budker}}, \ and\ \bibinfo {author}
  {\bibfnamefont {J.~R.}\ \bibnamefont {Maze}},\ }\href {\doibase
  10.1103/PhysRevB.97.094304} {\bibfield  {journal} {\bibinfo  {journal} {Phys.
  Rev. B}\ }\textbf {\bibinfo {volume} {97}},\ \bibinfo {pages} {094304}
  (\bibinfo {year} {2018})}\BibitemShut {NoStop}%
\bibitem [{\citenamefont {Gugler}\ \emph {et~al.}(2018)\citenamefont {Gugler},
  \citenamefont {Astner}, \citenamefont {Angerer}, \citenamefont
  {Schmiedmayer}, \citenamefont {Majer},\ and\ \citenamefont
  {Mohn}}]{GuglerT1T2018}%
  \BibitemOpen
  \bibfield  {author} {\bibinfo {author} {\bibfnamefont {J.}~\bibnamefont
  {Gugler}}, \bibinfo {author} {\bibfnamefont {T.}~\bibnamefont {Astner}},
  \bibinfo {author} {\bibfnamefont {A.}~\bibnamefont {Angerer}}, \bibinfo
  {author} {\bibfnamefont {J.}~\bibnamefont {Schmiedmayer}}, \bibinfo {author}
  {\bibfnamefont {J.}~\bibnamefont {Majer}}, \ and\ \bibinfo {author}
  {\bibfnamefont {P.}~\bibnamefont {Mohn}},\ }\href {\doibase
  10.1103/PhysRevB.98.214442} {\bibfield  {journal} {\bibinfo  {journal} {Phys.
  Rev. B}\ }\textbf {\bibinfo {volume} {98}},\ \bibinfo {pages} {214442}
  (\bibinfo {year} {2018})}\BibitemShut {NoStop}%
\bibitem [{\citenamefont {Iv\'ady}\ \emph {et~al.}(2015)\citenamefont
  {Iv\'ady}, \citenamefont {Sz\'asz}, \citenamefont {Falk}, \citenamefont
  {Klimov}, \citenamefont {Christle}, \citenamefont {Janz\'en}, \citenamefont
  {Abrikosov}, \citenamefont {Awschalom},\ and\ \citenamefont
  {Gali}}]{IvadyDNP2015}%
  \BibitemOpen
  \bibfield  {author} {\bibinfo {author} {\bibfnamefont {V.}~\bibnamefont
  {Iv\'ady}}, \bibinfo {author} {\bibfnamefont {K.}~\bibnamefont {Sz\'asz}},
  \bibinfo {author} {\bibfnamefont {A.~L.}\ \bibnamefont {Falk}}, \bibinfo
  {author} {\bibfnamefont {P.~V.}\ \bibnamefont {Klimov}}, \bibinfo {author}
  {\bibfnamefont {D.~J.}\ \bibnamefont {Christle}}, \bibinfo {author}
  {\bibfnamefont {E.}~\bibnamefont {Janz\'en}}, \bibinfo {author}
  {\bibfnamefont {I.~A.}\ \bibnamefont {Abrikosov}}, \bibinfo {author}
  {\bibfnamefont {D.~D.}\ \bibnamefont {Awschalom}}, \ and\ \bibinfo {author}
  {\bibfnamefont {A.}~\bibnamefont {Gali}},\ }\href {\doibase
  10.1103/PhysRevB.92.115206} {\bibfield  {journal} {\bibinfo  {journal} {Phys.
  Rev. B}\ }\textbf {\bibinfo {volume} {92}},\ \bibinfo {pages} {115206}
  (\bibinfo {year} {2015})}\BibitemShut {NoStop}%
\bibitem [{\citenamefont {Iv\'ady}\ \emph {et~al.}(2016)\citenamefont
  {Iv\'ady}, \citenamefont {Klimov}, \citenamefont {Miao}, \citenamefont
  {Falk}, \citenamefont {Christle}, \citenamefont {Sz\'asz}, \citenamefont
  {Abrikosov}, \citenamefont {Awschalom},\ and\ \citenamefont
  {Gali}}]{IvadyPRL2016}%
  \BibitemOpen
  \bibfield  {author} {\bibinfo {author} {\bibfnamefont {V.}~\bibnamefont
  {Iv\'ady}}, \bibinfo {author} {\bibfnamefont {P.~V.}\ \bibnamefont {Klimov}},
  \bibinfo {author} {\bibfnamefont {K.~C.}\ \bibnamefont {Miao}}, \bibinfo
  {author} {\bibfnamefont {A.~L.}\ \bibnamefont {Falk}}, \bibinfo {author}
  {\bibfnamefont {D.~J.}\ \bibnamefont {Christle}}, \bibinfo {author}
  {\bibfnamefont {K.}~\bibnamefont {Sz\'asz}}, \bibinfo {author} {\bibfnamefont
  {I.~A.}\ \bibnamefont {Abrikosov}}, \bibinfo {author} {\bibfnamefont {D.~D.}\
  \bibnamefont {Awschalom}}, \ and\ \bibinfo {author} {\bibfnamefont
  {A.}~\bibnamefont {Gali}},\ }\href {\doibase 10.1103/PhysRevLett.117.220503}
  {\bibfield  {journal} {\bibinfo  {journal} {Phys. Rev. Lett.}\ }\textbf
  {\bibinfo {volume} {117}},\ \bibinfo {pages} {220503} (\bibinfo {year}
  {2016})}\BibitemShut {NoStop}%
\bibitem [{\citenamefont {Wunderlich}\ \emph {et~al.}(2017)\citenamefont
  {Wunderlich}, \citenamefont {Kohlrautz}, \citenamefont {Abel}, \citenamefont
  {Haase},\ and\ \citenamefont {Meijer}}]{Wunderlich2017}%
  \BibitemOpen
  \bibfield  {author} {\bibinfo {author} {\bibfnamefont {R.}~\bibnamefont
  {Wunderlich}}, \bibinfo {author} {\bibfnamefont {J.}~\bibnamefont
  {Kohlrautz}}, \bibinfo {author} {\bibfnamefont {B.}~\bibnamefont {Abel}},
  \bibinfo {author} {\bibfnamefont {J.}~\bibnamefont {Haase}}, \ and\ \bibinfo
  {author} {\bibfnamefont {J.}~\bibnamefont {Meijer}},\ }\href {\doibase
  10.1103/PhysRevB.96.220407} {\bibfield  {journal} {\bibinfo  {journal} {Phys.
  Rev. B}\ }\textbf {\bibinfo {volume} {96}},\ \bibinfo {pages} {220407}
  (\bibinfo {year} {2017})}\BibitemShut {NoStop}%
\bibitem [{\citenamefont {Anishchik}\ and\ \citenamefont
  {Ivanov}(2019)}]{ANISHCHIK201967}%
  \BibitemOpen
  \bibfield  {author} {\bibinfo {author} {\bibfnamefont {S.}~\bibnamefont
  {Anishchik}}\ and\ \bibinfo {author} {\bibfnamefont {K.}~\bibnamefont
  {Ivanov}},\ }\href {\doibase https://doi.org/10.1016/j.jmr.2019.06.002}
  {\bibfield  {journal} {\bibinfo  {journal} {Journal of Magnetic Resonance}\
  }\textbf {\bibinfo {volume} {305}},\ \bibinfo {pages} {67 } (\bibinfo {year}
  {2019})}\BibitemShut {NoStop}%
\bibitem [{\citenamefont {Yang}\ \emph {et~al.}(2019)\citenamefont {Yang},
  \citenamefont {Wang}, \citenamefont {Tao}, \citenamefont {Yang},
  \citenamefont {Zhang}, \citenamefont {Ai},\ and\ \citenamefont
  {Deng}}]{Yang2019}%
  \BibitemOpen
  \bibfield  {author} {\bibinfo {author} {\bibfnamefont {Z.-S.}\ \bibnamefont
  {Yang}}, \bibinfo {author} {\bibfnamefont {Y.-X.}\ \bibnamefont {Wang}},
  \bibinfo {author} {\bibfnamefont {M.-J.}\ \bibnamefont {Tao}}, \bibinfo
  {author} {\bibfnamefont {W.}~\bibnamefont {Yang}}, \bibinfo {author}
  {\bibfnamefont {M.}~\bibnamefont {Zhang}}, \bibinfo {author} {\bibfnamefont
  {Q.}~\bibnamefont {Ai}}, \ and\ \bibinfo {author} {\bibfnamefont {F.-G.}\
  \bibnamefont {Deng}},\ }\href@noop {} {\bibfield  {journal} {\bibinfo
  {journal} {arXiv:1804.01008}\ } (\bibinfo {year} {2019})}\BibitemShut
  {NoStop}%
\bibitem [{\citenamefont {Al-Hassanieh}\ \emph {et~al.}(2006)\citenamefont
  {Al-Hassanieh}, \citenamefont {Dobrovitski}, \citenamefont {Dagotto},\ and\
  \citenamefont {Harmon}}]{AlHassaniehPRL2006}%
  \BibitemOpen
  \bibfield  {author} {\bibinfo {author} {\bibfnamefont {K.~A.}\ \bibnamefont
  {Al-Hassanieh}}, \bibinfo {author} {\bibfnamefont {V.~V.}\ \bibnamefont
  {Dobrovitski}}, \bibinfo {author} {\bibfnamefont {E.}~\bibnamefont
  {Dagotto}}, \ and\ \bibinfo {author} {\bibfnamefont {B.~N.}\ \bibnamefont
  {Harmon}},\ }\href {\doibase 10.1103/PhysRevLett.97.037204} {\bibfield
  {journal} {\bibinfo  {journal} {Phys. Rev. Lett.}\ }\textbf {\bibinfo
  {volume} {97}},\ \bibinfo {pages} {037204} (\bibinfo {year}
  {2006})}\BibitemShut {NoStop}%
\bibitem [{\citenamefont {Sinitsyn}\ \emph {et~al.}(2012)\citenamefont
  {Sinitsyn}, \citenamefont {Li}, \citenamefont {Crooker}, \citenamefont
  {Saxena},\ and\ \citenamefont {Smith}}]{SinitsynPRL2012}%
  \BibitemOpen
  \bibfield  {author} {\bibinfo {author} {\bibfnamefont {N.~A.}\ \bibnamefont
  {Sinitsyn}}, \bibinfo {author} {\bibfnamefont {Y.}~\bibnamefont {Li}},
  \bibinfo {author} {\bibfnamefont {S.~A.}\ \bibnamefont {Crooker}}, \bibinfo
  {author} {\bibfnamefont {A.}~\bibnamefont {Saxena}}, \ and\ \bibinfo {author}
  {\bibfnamefont {D.~L.}\ \bibnamefont {Smith}},\ }\href {\doibase
  10.1103/PhysRevLett.109.166605} {\bibfield  {journal} {\bibinfo  {journal}
  {Phys. Rev. Lett.}\ }\textbf {\bibinfo {volume} {109}},\ \bibinfo {pages}
  {166605} (\bibinfo {year} {2012})}\BibitemShut {NoStop}%
\bibitem [{\citenamefont {Udvarhelyi}\ \emph {et~al.}(2018)\citenamefont
  {Udvarhelyi}, \citenamefont {Shkolnikov}, \citenamefont {Gali}, \citenamefont
  {Burkard},\ and\ \citenamefont {P\'alyi}}]{UdvarhelyiStrain2018}%
  \BibitemOpen
  \bibfield  {author} {\bibinfo {author} {\bibfnamefont {P.}~\bibnamefont
  {Udvarhelyi}}, \bibinfo {author} {\bibfnamefont {V.~O.}\ \bibnamefont
  {Shkolnikov}}, \bibinfo {author} {\bibfnamefont {A.}~\bibnamefont {Gali}},
  \bibinfo {author} {\bibfnamefont {G.}~\bibnamefont {Burkard}}, \ and\
  \bibinfo {author} {\bibfnamefont {A.}~\bibnamefont {P\'alyi}},\ }\href
  {\doibase 10.1103/PhysRevB.98.075201} {\bibfield  {journal} {\bibinfo
  {journal} {Phys. Rev. B}\ }\textbf {\bibinfo {volume} {98}},\ \bibinfo
  {pages} {075201} (\bibinfo {year} {2018})}\BibitemShut {NoStop}%
\bibitem [{\citenamefont {Felton}\ \emph {et~al.}(2008)\citenamefont {Felton},
  \citenamefont {Edmonds}, \citenamefont {Newton}, \citenamefont {Martineau},
  \citenamefont {Fisher},\ and\ \citenamefont {Twitchen}}]{Felton2008}%
  \BibitemOpen
  \bibfield  {author} {\bibinfo {author} {\bibfnamefont {S.}~\bibnamefont
  {Felton}}, \bibinfo {author} {\bibfnamefont {A.~M.}\ \bibnamefont {Edmonds}},
  \bibinfo {author} {\bibfnamefont {M.~E.}\ \bibnamefont {Newton}}, \bibinfo
  {author} {\bibfnamefont {P.~M.}\ \bibnamefont {Martineau}}, \bibinfo {author}
  {\bibfnamefont {D.}~\bibnamefont {Fisher}}, \ and\ \bibinfo {author}
  {\bibfnamefont {D.~J.}\ \bibnamefont {Twitchen}},\ }\href {\doibase
  10.1103/PhysRevB.77.081201} {\bibfield  {journal} {\bibinfo  {journal} {Phys.
  Rev. B}\ }\textbf {\bibinfo {volume} {77}},\ \bibinfo {pages} {081201}
  (\bibinfo {year} {2008})}\BibitemShut {NoStop}%
\bibitem [{\citenamefont {Heyd}\ \emph {et~al.}(2003)\citenamefont {Heyd},
  \citenamefont {Scuseria},\ and\ \citenamefont {Ernzerhof}}]{HSE03}%
  \BibitemOpen
  \bibfield  {author} {\bibinfo {author} {\bibfnamefont {J.}~\bibnamefont
  {Heyd}}, \bibinfo {author} {\bibfnamefont {G.~E.}\ \bibnamefont {Scuseria}},
  \ and\ \bibinfo {author} {\bibfnamefont {M.}~\bibnamefont {Ernzerhof}},\
  }\href {\doibase 10.1063/1.1564060} {\bibfield  {journal} {\bibinfo
  {journal} {J. Chem. Phys.}\ }\textbf {\bibinfo {volume} {118}},\ \bibinfo
  {pages} {8207} (\bibinfo {year} {2003})}\BibitemShut {NoStop}%
\bibitem [{\citenamefont {Bl\"ochl}(1994)}]{PAW}%
  \BibitemOpen
  \bibfield  {author} {\bibinfo {author} {\bibfnamefont {P.~E.}\ \bibnamefont
  {Bl\"ochl}},\ }\href {\doibase 10.1103/PhysRevB.50.17953} {\bibfield
  {journal} {\bibinfo  {journal} {Phys. Rev. B}\ }\textbf {\bibinfo {volume}
  {50}},\ \bibinfo {pages} {17953} (\bibinfo {year} {1994})}\BibitemShut
  {NoStop}%
\bibitem [{\citenamefont {Kresse}\ and\ \citenamefont
  {Furthm\"uller}(1996)}]{VASP2}%
  \BibitemOpen
  \bibfield  {author} {\bibinfo {author} {\bibfnamefont {G.}~\bibnamefont
  {Kresse}}\ and\ \bibinfo {author} {\bibfnamefont {J.}~\bibnamefont
  {Furthm\"uller}},\ }\href {\doibase 10.1103/PhysRevB.54.11169} {\bibfield
  {journal} {\bibinfo  {journal} {Phys. Rev. B}\ }\textbf {\bibinfo {volume}
  {54}},\ \bibinfo {pages} {11169} (\bibinfo {year} {1996})}\BibitemShut
  {NoStop}%
\bibitem [{\citenamefont {Sz\'asz}\ \emph {et~al.}(2013)\citenamefont
  {Sz\'asz}, \citenamefont {Hornos}, \citenamefont {Marsman},\ and\
  \citenamefont {Gali}}]{Szasz13}%
  \BibitemOpen
  \bibfield  {author} {\bibinfo {author} {\bibfnamefont {K.}~\bibnamefont
  {Sz\'asz}}, \bibinfo {author} {\bibfnamefont {T.}~\bibnamefont {Hornos}},
  \bibinfo {author} {\bibfnamefont {M.}~\bibnamefont {Marsman}}, \ and\
  \bibinfo {author} {\bibfnamefont {A.}~\bibnamefont {Gali}},\ }\href {\doibase
  10.1103/PhysRevB.88.075202} {\bibfield  {journal} {\bibinfo  {journal} {Phys.
  Rev. B}\ }\textbf {\bibinfo {volume} {88}},\ \bibinfo {pages} {075202}
  (\bibinfo {year} {2013})}\BibitemShut {NoStop}%
\bibitem [{\citenamefont {Davidsson}\ \emph {et~al.}(2018)\citenamefont
  {Davidsson}, \citenamefont {Iv{\'{a}}dy}, \citenamefont {Armiento},
  \citenamefont {Son}, \citenamefont {Gali},\ and\ \citenamefont
  {Abrikosov}}]{Davidsson2018}%
  \BibitemOpen
  \bibfield  {author} {\bibinfo {author} {\bibfnamefont {J.}~\bibnamefont
  {Davidsson}}, \bibinfo {author} {\bibfnamefont {V.}~\bibnamefont
  {Iv{\'{a}}dy}}, \bibinfo {author} {\bibfnamefont {R.}~\bibnamefont
  {Armiento}}, \bibinfo {author} {\bibfnamefont {N.~T.}\ \bibnamefont {Son}},
  \bibinfo {author} {\bibfnamefont {A.}~\bibnamefont {Gali}}, \ and\ \bibinfo
  {author} {\bibfnamefont {I.~A.}\ \bibnamefont {Abrikosov}},\ }\href {\doibase
  10.1088/1367-2630/aaa752} {\bibfield  {journal} {\bibinfo  {journal} {New
  Journal of Physics}\ }\textbf {\bibinfo {volume} {20}},\ \bibinfo {pages}
  {023035} (\bibinfo {year} {2018})}\BibitemShut {NoStop}%
\bibitem [{\citenamefont {Iv{\'a}dy}\ \emph {et~al.}(2018)\citenamefont
  {Iv{\'a}dy}, \citenamefont {Abrikosov},\ and\ \citenamefont
  {Gali}}]{IvadyNPJ2018}%
  \BibitemOpen
  \bibfield  {author} {\bibinfo {author} {\bibfnamefont {V.}~\bibnamefont
  {Iv{\'a}dy}}, \bibinfo {author} {\bibfnamefont {I.~A.}\ \bibnamefont
  {Abrikosov}}, \ and\ \bibinfo {author} {\bibfnamefont {A.}~\bibnamefont
  {Gali}},\ }\href {\doibase 10.1038/s41524-018-0132-5} {\bibfield  {journal}
  {\bibinfo  {journal} {npj Computational Materials}\ }\textbf {\bibinfo
  {volume} {4}},\ \bibinfo {pages} {76} (\bibinfo {year} {2018})}\BibitemShut
  {NoStop}%
\bibitem [{\citenamefont {Jarmola}\ \emph {et~al.}(2012)\citenamefont
  {Jarmola}, \citenamefont {Acosta}, \citenamefont {Jensen}, \citenamefont
  {Chemerisov},\ and\ \citenamefont {Budker}}]{Jarmola2012}%
  \BibitemOpen
  \bibfield  {author} {\bibinfo {author} {\bibfnamefont {A.}~\bibnamefont
  {Jarmola}}, \bibinfo {author} {\bibfnamefont {V.~M.}\ \bibnamefont {Acosta}},
  \bibinfo {author} {\bibfnamefont {K.}~\bibnamefont {Jensen}}, \bibinfo
  {author} {\bibfnamefont {S.}~\bibnamefont {Chemerisov}}, \ and\ \bibinfo
  {author} {\bibfnamefont {D.}~\bibnamefont {Budker}},\ }\href {\doibase
  10.1103/PhysRevLett.108.197601} {\bibfield  {journal} {\bibinfo  {journal}
  {Phys. Rev. Lett.}\ }\textbf {\bibinfo {volume} {108}},\ \bibinfo {pages}
  {197601} (\bibinfo {year} {2012})}\BibitemShut {NoStop}%
\bibitem [{\citenamefont {Wood}\ \emph {et~al.}(2016)\citenamefont {Wood},
  \citenamefont {Broadway}, \citenamefont {Hall}, \citenamefont {Stacey},
  \citenamefont {Simpson}, \citenamefont {Tetienne},\ and\ \citenamefont
  {Hollenberg}}]{Wood2016}%
  \BibitemOpen
  \bibfield  {author} {\bibinfo {author} {\bibfnamefont {J.~D.~A.}\
  \bibnamefont {Wood}}, \bibinfo {author} {\bibfnamefont {D.~A.}\ \bibnamefont
  {Broadway}}, \bibinfo {author} {\bibfnamefont {L.~T.}\ \bibnamefont {Hall}},
  \bibinfo {author} {\bibfnamefont {A.}~\bibnamefont {Stacey}}, \bibinfo
  {author} {\bibfnamefont {D.~A.}\ \bibnamefont {Simpson}}, \bibinfo {author}
  {\bibfnamefont {J.-P.}\ \bibnamefont {Tetienne}}, \ and\ \bibinfo {author}
  {\bibfnamefont {L.~C.~L.}\ \bibnamefont {Hollenberg}},\ }\href {\doibase
  10.1103/PhysRevB.94.155402} {\bibfield  {journal} {\bibinfo  {journal} {Phys.
  Rev. B}\ }\textbf {\bibinfo {volume} {94}},\ \bibinfo {pages} {155402}
  (\bibinfo {year} {2016})}\BibitemShut {NoStop}%
\bibitem [{\citenamefont {Armstrong}\ \emph {et~al.}(2010)\citenamefont
  {Armstrong}, \citenamefont {Rogers}, \citenamefont {McMurtrie},\ and\
  \citenamefont {Manson}}]{Armstrong2010}%
  \BibitemOpen
  \bibfield  {author} {\bibinfo {author} {\bibfnamefont {S.}~\bibnamefont
  {Armstrong}}, \bibinfo {author} {\bibfnamefont {L.~J.}\ \bibnamefont
  {Rogers}}, \bibinfo {author} {\bibfnamefont {R.~L.}\ \bibnamefont
  {McMurtrie}}, \ and\ \bibinfo {author} {\bibfnamefont {N.~B.}\ \bibnamefont
  {Manson}},\ }\href {\doibase https://doi.org/10.1016/j.phpro.2010.01.223}
  {\bibfield  {journal} {\bibinfo  {journal} {Physics Procedia}\ }\textbf
  {\bibinfo {volume} {3}},\ \bibinfo {pages} {1569 } (\bibinfo {year}
  {2010})},\ \bibinfo {note} {proceedings of the Tenth International Meeting on
  Hole Burning, Single Molecule and Related Spectroscopies: Science and
  Applications-HBSM 2009}\BibitemShut {NoStop}%
\bibitem [{\citenamefont {Hall}\ \emph {et~al.}(2016)\citenamefont {Hall},
  \citenamefont {Kehayias}, \citenamefont {Simpson}, \citenamefont {Jarmola},
  \citenamefont {Stacey}, \citenamefont {Budker},\ and\ \citenamefont
  {Hollenberg}}]{Hall2016}%
  \BibitemOpen
  \bibfield  {author} {\bibinfo {author} {\bibfnamefont {L.~T.}\ \bibnamefont
  {Hall}}, \bibinfo {author} {\bibfnamefont {P.}~\bibnamefont {Kehayias}},
  \bibinfo {author} {\bibfnamefont {D.~A.}\ \bibnamefont {Simpson}}, \bibinfo
  {author} {\bibfnamefont {A.}~\bibnamefont {Jarmola}}, \bibinfo {author}
  {\bibfnamefont {A.}~\bibnamefont {Stacey}}, \bibinfo {author} {\bibfnamefont
  {D.}~\bibnamefont {Budker}}, \ and\ \bibinfo {author} {\bibfnamefont
  {L.~C.~L.}\ \bibnamefont {Hollenberg}},\ }\href {\doibase
  10.1038/ncomms10211} {\bibfield  {journal} {\bibinfo  {journal} {Nature
  Communications}\ }\textbf {\bibinfo {volume} {7}},\ \bibinfo {pages} {10211}
  (\bibinfo {year} {2016})}\BibitemShut {NoStop}%
\bibitem [{\citenamefont {Wickenbrock}\ \emph {et~al.}(2016)\citenamefont
  {Wickenbrock}, \citenamefont {Zheng}, \citenamefont {Bougas}, \citenamefont
  {Leefer}, \citenamefont {Afach}, \citenamefont {Jarmola}, \citenamefont
  {Acosta},\ and\ \citenamefont {Budker}}]{Wickenbrock2016}%
  \BibitemOpen
  \bibfield  {author} {\bibinfo {author} {\bibfnamefont {A.}~\bibnamefont
  {Wickenbrock}}, \bibinfo {author} {\bibfnamefont {H.}~\bibnamefont {Zheng}},
  \bibinfo {author} {\bibfnamefont {L.}~\bibnamefont {Bougas}}, \bibinfo
  {author} {\bibfnamefont {N.}~\bibnamefont {Leefer}}, \bibinfo {author}
  {\bibfnamefont {S.}~\bibnamefont {Afach}}, \bibinfo {author} {\bibfnamefont
  {A.}~\bibnamefont {Jarmola}}, \bibinfo {author} {\bibfnamefont {V.~M.}\
  \bibnamefont {Acosta}}, \ and\ \bibinfo {author} {\bibfnamefont
  {D.}~\bibnamefont {Budker}},\ }\href {\doibase 10.1063/1.4960171} {\bibfield
  {journal} {\bibinfo  {journal} {Applied Physics Letters}\ }\textbf {\bibinfo
  {volume} {109}},\ \bibinfo {pages} {053505} (\bibinfo {year}
  {2016})}\BibitemShut {NoStop}%
\bibitem [{\citenamefont {Fischer}\ \emph {et~al.}(2013)\citenamefont
  {Fischer}, \citenamefont {Bretschneider}, \citenamefont {London},
  \citenamefont {Budker}, \citenamefont {Gershoni},\ and\ \citenamefont
  {Frydman}}]{FischerPRL2013}%
  \BibitemOpen
  \bibfield  {author} {\bibinfo {author} {\bibfnamefont {R.}~\bibnamefont
  {Fischer}}, \bibinfo {author} {\bibfnamefont {C.~O.}\ \bibnamefont
  {Bretschneider}}, \bibinfo {author} {\bibfnamefont {P.}~\bibnamefont
  {London}}, \bibinfo {author} {\bibfnamefont {D.}~\bibnamefont {Budker}},
  \bibinfo {author} {\bibfnamefont {D.}~\bibnamefont {Gershoni}}, \ and\
  \bibinfo {author} {\bibfnamefont {L.}~\bibnamefont {Frydman}},\ }\href
  {\doibase 10.1103/PhysRevLett.111.057601} {\bibfield  {journal} {\bibinfo
  {journal} {Phys. Rev. Lett.}\ }\textbf {\bibinfo {volume} {111}},\ \bibinfo
  {pages} {057601} (\bibinfo {year} {2013})}\BibitemShut {NoStop}%
\end{thebibliography}

%

\end{document}